\documentclass[12pt,a4paper]{JHEP3}
\usepackage{amsmath}
\usepackage{amssymb}
\usepackage{mathrsfs}
\usepackage{graphicx}
\usepackage{subfigure}
\usepackage{multirow}
\usepackage{epsfig}

\def\stop{{\tilde t}}

\def\NO{{\tilde\chi^0_1}}
\def\MeV{{\textrm{MeV}/\textrm{c}^{2}}}
\def\GeV{{\textrm{GeV}/\textrm{c}^{2}}}
\def\TeV{{\textrm{TeV}/\textrm{c}^{2}}}

\def\GeVm{{\textrm{GeV}/\textrm{c}}}

\def\lapprox{\ensuremath{\sim\kern-1em\raise 0.65ex\hbox{$<$}}}%  Or use \lsim
\def\rapprox{\ensuremath{\sim\kern-1em\raise 0.65ex\hbox{$>$}}}%  and \rsim.

\title{Long-lived stops in MSSM scenarios with a neutralino LSP}

\author{M. Johansen\footnote{marianne.johansen@fysik.su.se}, J. Edsj\"o, S. Hellman, D. Milstead \\ \\
  \small{Department of Physics and } \\
 \small{the Oskar Klein Centre for Cosmoparticle Physics}\\
   \small{Stockholm University}\\
   \small{SE - 106 91 Stockholm}}

\abstract{This work investigates the possibility of a long-lived stop squark in supersymmetric models with the neutralino as the lightest supersymmetric particle (LSP). We study the implications of meta-stable stops on the sparticle mass spectra and the dark matter density. We find that in order to obtain a sufficiently long stop lifetime so as to be observable as a stable $R$-hadron at an LHC experiment, we need to fine tune the mass degeneracy
  between the stop and the LSP considerably. This increases the stop-neutralino co-anihilation cross section, leaving the neutralino relic density lower than what is expected from the WMAP results for stop masses $\lapprox 1.5$ TeV/$c^2$. However, if such scenarios are realised in nature we demonstrate that the long-lived stops will be produced at the LHC and that stop-based $R$-hadrons with masses up to 1 $\TeV$ can be detected after one year of running at design luminosity.}

\begin{document}

\date{}
\maketitle

\section{Introduction}
The well known problems of the lack of naturalness of the Standard Model and its inability to explain dark matter have led to an expectation of hitherto unobserved particles and physics processes being manifest at TeV energy scales\cite{Gabadadze:2003jk}. A number of candidate phenomenological extensions of the Standard Model have been proposed although, as yet, there is no experimental evidence to either falsify or confirm the correctness of any of these models. It is therefore prudent to consider a range of observables which could be used to discover and quantify any exotic physics signals seen at the TeV scale. Such observables include anomalous rates of jet and missing transverse energy production, and the observation of exotic stable\footnote{In this paper, the terms stable  and long-lived particle are
taken to imply that the particle will not decay during its traversal of a typical detector at a collider unless stated otherwise.}
  massive interacting particles (SMPs), which is the subject of this paper. The detection of SMPs is experimentally challenging owing to their slow speeds and interactions in matter. In this work, it is shown how a stable stop can arise in supersymmetric scenarios\cite{Bergstrom:1995cz} in which the neutralino is the lightest supersymmetric particle (LSP). The consistency of such scenarios with cosmological constraints is also studied.  Stable stops would manifest themselves in an experiment as bound hadronic states (termed $R$-hadrons). Using the {\sc Pythia}\cite{pythia} generator, the signatures of MSSM-8 models which give rise to $R$-hadrons at the LHC  are investigated.

\par Long-lived stops are predicted in many theoretical models\cite{Fairbairn:2006gg,Raklev:2009mg}. In the theory of electroweak baryogenesis \cite{ewbg1,ewbg2,ewbg3} a long lived stop is predicted in the so-called coannhilation region\cite{stopEwbg,are,stopLH}. The stop lifetime in these models does not typically exceed $10^{-15}$ s, leaving no possibility of  them traversing any substantial part of a detector at a high-energy collider experiment. Supersymmetry (SUSY) models with a gravitino LSP\cite{gravitinoLSP1,gravitinoLSP2,gravitinoLSP3} can also give rise to a long-lived stop if it is the next-to-lightest supersymmetric particle (NLSP). Due to the smallness of the gravitational coupling these stops can have a macroscopic lifetime\cite{gravitinoStop1}. Lifetimes with similar values can be obtained from models where the right-handed sneutrino is the LSP\cite{sneutrinoStop}, since the coupling is proportional to the neutralino Yukawa coupling. A stop NLSP can be obtained when non-universal third family sfermion masses are introduced. It should be noted, however, that none of these models combine a neutralino LSP with a long-lived stop as is done in the models studied here.

\par There have been a number of searches\cite{Fairbairn:2006gg} for $R$-hadrons at colliders.  The most recent searches were performed with data from LEP\cite{delphiStop,alephRhadron} and the Tevatron\cite{lepRhad,tevRhad}. Guided by these results, we consider the possibility of detecting at the LHC stable stops with masses in excess of the excluded lower limit $\sim 249$ GeV/$c^2$\cite{cdfStops}. This work complements studies which have been made of the discovery potential of $R$-hadrons at the LHC\cite{aliceRhad,atlasCSC,aafke,paRhad} and at cosmic ray facilities such as the Pierre Auger\cite{pierreAuger} and IceCube\cite{Hewett:2004nw} experiments.

\section{Long-lived stops in the MSSM-8}
\begin{figure}[h!]
\centering
\subfigure[ ]{
\label{fig:hO_xsec}
\includegraphics[width=0.30\textwidth]{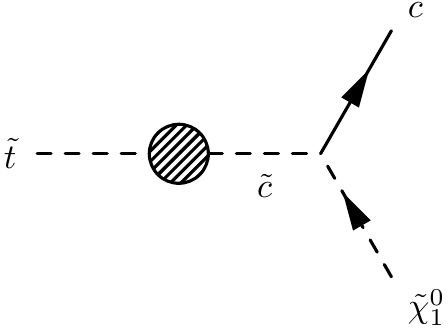}
}
\subfigure[ ]{
\label{fig:hO_xsec}
\includegraphics[width=0.30\textwidth]{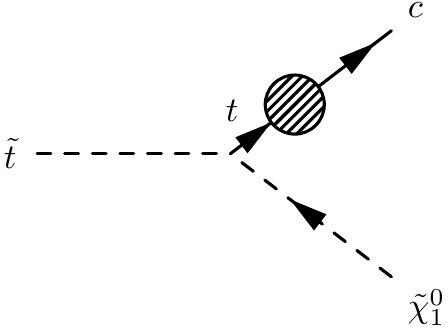}
}
\subfigure[]{
\label{fig:hO_xsec}
\includegraphics[width=0.30\textwidth]{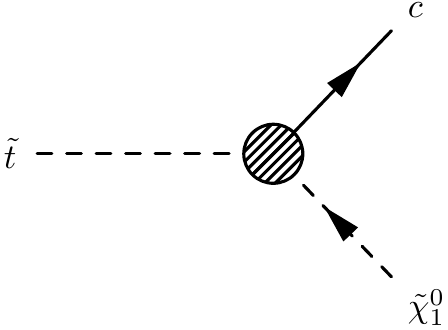}
}

\caption{\label{fig:stopDecay} One loop stop decays\protect \cite{stopDecay}. }
\end{figure}

A long-lived stop in models with a neutralino LSP can arise from a number of different parameter settings in the MSSM\cite{mssm}. In such scenarios the stop is the NLSP and can be long-lived if it is very close in mass to the LSP. In that case only a very limited number of decay modes are open. In fact if the mass difference between the stop and the LSP is less than the $W$-mass, only the suppressed modes:
\begin{eqnarray}
\label{eq:stopCdecay}\stop & \rightarrow & c\NO\\
\label{eq:stopUdecay}	& \rightarrow & u\NO\\
	& \rightarrow  & bf\bar{f'}\NO,
\end{eqnarray}
are allowed. Figure \ref{fig:stopDecay} shows  the Feynman graphs for the decay $\stop \rightarrow c\NO$.
\par In most phenomenological SUSY scenarios (such as MSSM-7\cite{Bergstrom:1995cz} or mSUGRA/    CMSSM\cite{mSugra}), for simplicity one assumes a degeneracy in the sfermion sector (either at the Grand Unified Theory (GUT) scale, or at the electroweak scale). As we are here mostly interested in models with a light stop, but not necessarily with all sfermions being light, we shall break this degeneracy by introducing an additional mass parameter for the third generation of squarks. Our phenomenological MSSM-8 model then has the following free parameters,
\begin{itemize}
\item the top trilinear coupling $A_{t}$
\item the bottom trilinear coupling $A_{b}$
\item the ratio of the vacuum expectation value of the two Higgs doublets $\tan \beta = \frac{v_{2}}{v_{1}}$
\item the pseudoscalar Higgs mass $m_{A}$
\item the $\mu$ parameter
\item the common SU(2) gaugino mass $M_{2}$
\item the first and second generation sfermion mass ${m}_{0}$
\item the third generation sfermion mass ${m}_{03}$
\end{itemize}
This is the same model as the MSSM-7 model in Ref. \cite{Bergstrom:1995cz}, but with the extra parameter ${m}_{03}$ which replaces  ${m}_{0}$ in the (3,3)-component of the squark matriceses $M_{Q}$, $M_{U}$ and $M_{D}$ (see Ref. \cite{Bergstrom:1995cz} for more details). The non-universal sfermion mass scenario, introduced to separate the third generation of squarks from the first and the second, allows for  the construction of models in which the stop and neutralino are close in mass. Note that a similar mass spectrum can be obtained in mSUGRA/CMSSM with a non-zero trilinear coupling at the GUT scale although here we prefer to work in the more general MSSM-8 model. Furthermore, we only need one gaugino mass parameter as we make the usual GUT assumptions for the gaugino mass parameters:
\begin{equation}
M_{1}\sim 0.5 M_{2}
\end{equation}
\begin{equation}
M_{2}\sim 0.3 M_{3}.
\end{equation}

The degeneracy between the stop and the neutralino masses influences the neutralino relic density via so-called coannihilations \cite{coann}. Hence, if we let the stop be arbitrarily close in mass to the neutralino we would typically have too much coannihilation, leading to an effective annihilation cross section which is too large and a relic density which is too low. This condition is especially severe for lighter neutralinos, where the neutralino-neutralino annihilation cross section is fairly large already without coannihilations. Therefore, to respect the relic density constraint, we cannot allow mass differences between the stop and the neutralino which are too small at low masses, whereas at higher masses (TeV scale), we can let the stop be more degenerate with the neutralino.

\section{Model scan}

\par Models with a tiny mass difference between the stop and the LSP were generated using DarkSUSY\cite{darkSusy}. The models were checked against current experimental constraints, as implemented in DarkSUSY:
\begin{itemize}
\item masses of superpartners \cite{pdg}
\item masses of Higgs particles, checked with HiggsBounds 1.0.3 \cite{Bechtle:2008jh}
\item the invisible width of the $Z$ boson \cite{pdg}
\item the $\rho$-parameter \cite{pdg}
\item the rare decay $b \rightarrow s \gamma$, where we assume that the standard model and SUSY contributions gives a decay branching fraction in the range $2.71 \times 10^{-4} < Br(b \rightarrow s \gamma) < 4.39 \times 10^{-4}$, in agreement with the experimental world average \cite{Barberio:2007cr} and allowing for theoretical and experimental errors
\item the magnetic moment of the muon, where we assume that $-2.7\times 10^{-10} < a_\mu = (g-2)_\mu/2 < 49 \times 10^{-10}$, in agreement with the experimental measurement \cite{Bennett:2006fi} and allowing for theoretical and experimental errors
\end{itemize}
For more details on these limits, see Ref. \cite{darkSusy}.

In total, we have generated many thousands of different models with different masses and compositions. The focus has been on models with a small mass difference between the neutralino and the lightest stop. We have also chosen to apply the relic density constraint on $\Omega_{\tilde{\chi}} h^{2}$ with various degrees of hardness,
\begin{itemize}
\item[1.] a hard requirement $0.1099 \pm 2 \cdot 0.0062$, corresponding to the five year WMAP result $\pm 2\sigma$\cite{wmap}.
\item[2.] a soft requirement  $0.05<\Omega_{\tilde{\chi}}  h^{2} < 0.2$, corresponding to $\Omega_{\tilde{\chi}}  h^{2}=0.1099^{+14 \sigma}_{-10\sigma}$.
\item[3.] no requirement on the value of  $\Omega_{\tilde{\chi}} {h}^{2}$ \footnote{In practice all our models have a value of $\Omega_{\tilde{\chi}} {h}^{2}$ well below the five year WMAP result.}.
\end{itemize}

For practical reasons, we have selected a set of representative benchmark models with different neutralino masses by binning our models in 100 logarithmic intervals between $10$ $\GeV$ and $10$ $\TeV$ in neutralino mass. In each bin, we have selected the model passing all the constraints given above (for the different scenarios of the relic density) that has the smallest relative mass difference between the LSP and the lightest stop i.e.\ where $(m_{\tilde{t}} -m_{\tilde{\chi}})/m_{\tilde{\chi}}$ is the lowest. This was done to ensure long stop lifetimes.

The parameters and masses of the chosen models are written to SUSY Les Houches Accord 2 (SLHA)\cite{slha2} files. These are used as inputs to {\sc SDECAY}\cite{sdecay} which calculates the decay rate and width of the SUSY particles using loop corrections such as those shown in Figure \ref{fig:stopDecay}. It is important for the decays in Eq.~(\ref{eq:stopCdecay}) and (\ref{eq:stopUdecay}) that both the $c$ and $u$ masses are set to zero by default. The $c$-quark mass, $m_{c}=1.21\, \GeV$,  is therefore added to {\sc SDECAY} by hand to avoid overestimating the stop decay rate contribution from Eq.~(\ref{eq:stopCdecay}). The $u$-quark mass is left as zero, which is a conservative choice; any variations of the mass would lead to a longer stop lifetime. SLHA output from {\sc SDECAY}  contains the decay branching ratios and widths for all sparticles. From the width, the stop lifetime can be obtained, and models that provide stops with a lifetime longer than 100 ns are then used as input to the event generator {\sc Pythia}\cite{pythia}, version 6.4.  The {\sc Pythia} program generates events and calculates both the total MSSM cross section and the cross section for each sub-process. Kinematical event distributions such as momentum, angles and particle multiplicities are also provided.
\par Many of our benchmark models depend sensitively on mass differences and hence on the input parameters. As the relic density depends exponentially on the mass difference between the lightest neutralino and the lightest stop in the coannihilation region, it is important to have sufficient numerical precision for this calculation. We have therefore generated all models and calculated accelerator constraints and the relic density in double precision. When we have calculated the decay rates with {\sc SDECAY},  we have used SLHA2 to transfer our model parameters, which does not allow for double precision. Hence a small numerical error occurs. For a few models with extremely small mass differences, this numerical error was found to be significant, and those models were excluded from our set of models. For the remaining models, the numerical error is small enough not to be of any problem for the current study.
\begin{figure}[t!]

\subfigure[ $\tau_{\tilde{t}}$ vs $m_{\tilde{t}}-m_{\tilde{\chi}}$.]{
\label{fig:hO_lt}
\includegraphics[width=
0.45\textwidth, height= 5 cm]{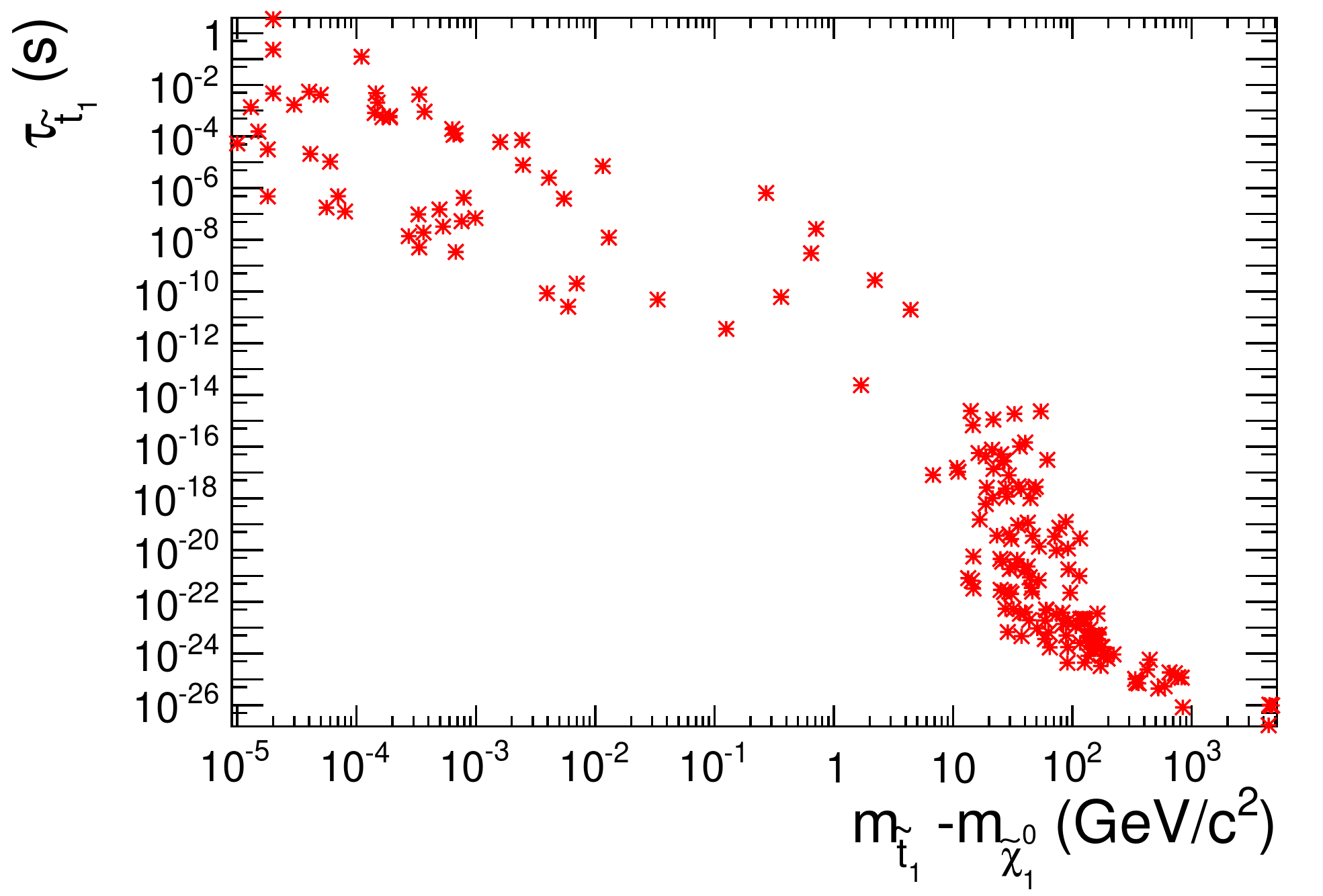}
}
\hskip 1 cm
\subfigure[$m_{\tilde{t}}-m_{\tilde{\chi}}$ vs $m_{\tilde{\chi}}$.]{
\label{fig:hO_m}
\includegraphics[width=
0.45\textwidth,height= 5 cm]{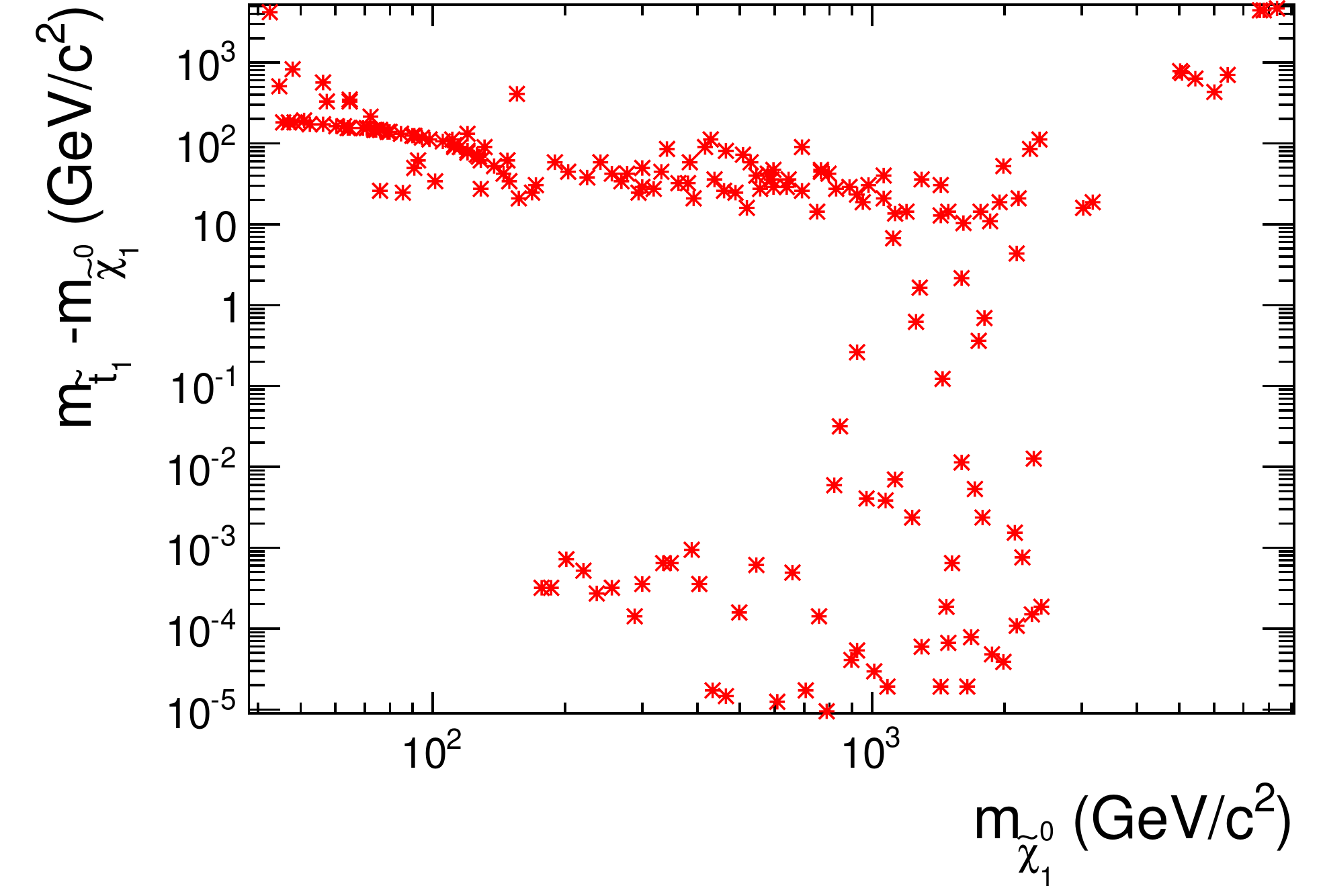}
}

\caption{\label{fig:lifetime} The left plot shows plots of the stop lifetime as a function of the mass difference between the stop and neutralino for all models. The right plot shows plots of the stop/neutralino mass difference versus the neutralino mass.}
\end{figure}

\begin{figure}[htbp]
\subfigure[$\tau_{\tilde{t}}$ vs $m_{\tilde{t}}-m_{\tilde{\chi}}$ for  $\tau_{\tilde{t}}>100$ ns, hard $\Omega_{\tilde{\chi}} $ cut.]{
\label{fig:hO_lt_stable}
\includegraphics[width=
0.45\textwidth, height= 4.7 cm]{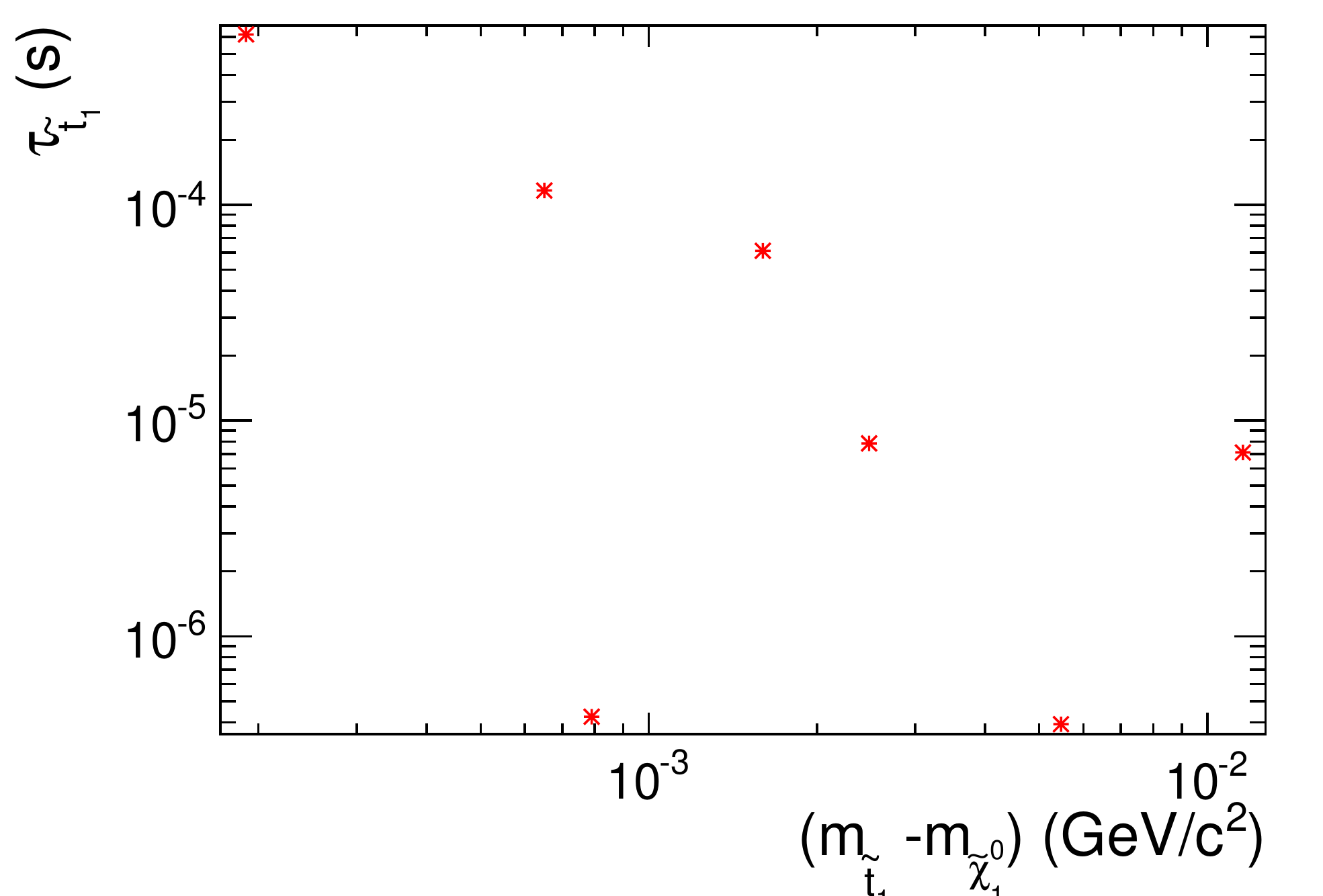}
}
\hskip 1 cm
\subfigure[$m_{\tilde{\chi}}$ vs $m_{\tilde{t}}$ for  $\tau_{\tilde{t}}>100$ ns, hard $\Omega_{\tilde{\chi}} $ cut.]{
\label{fig:hO_m_stable}
\includegraphics[width=
0.45\textwidth, height= 4.7 cm]{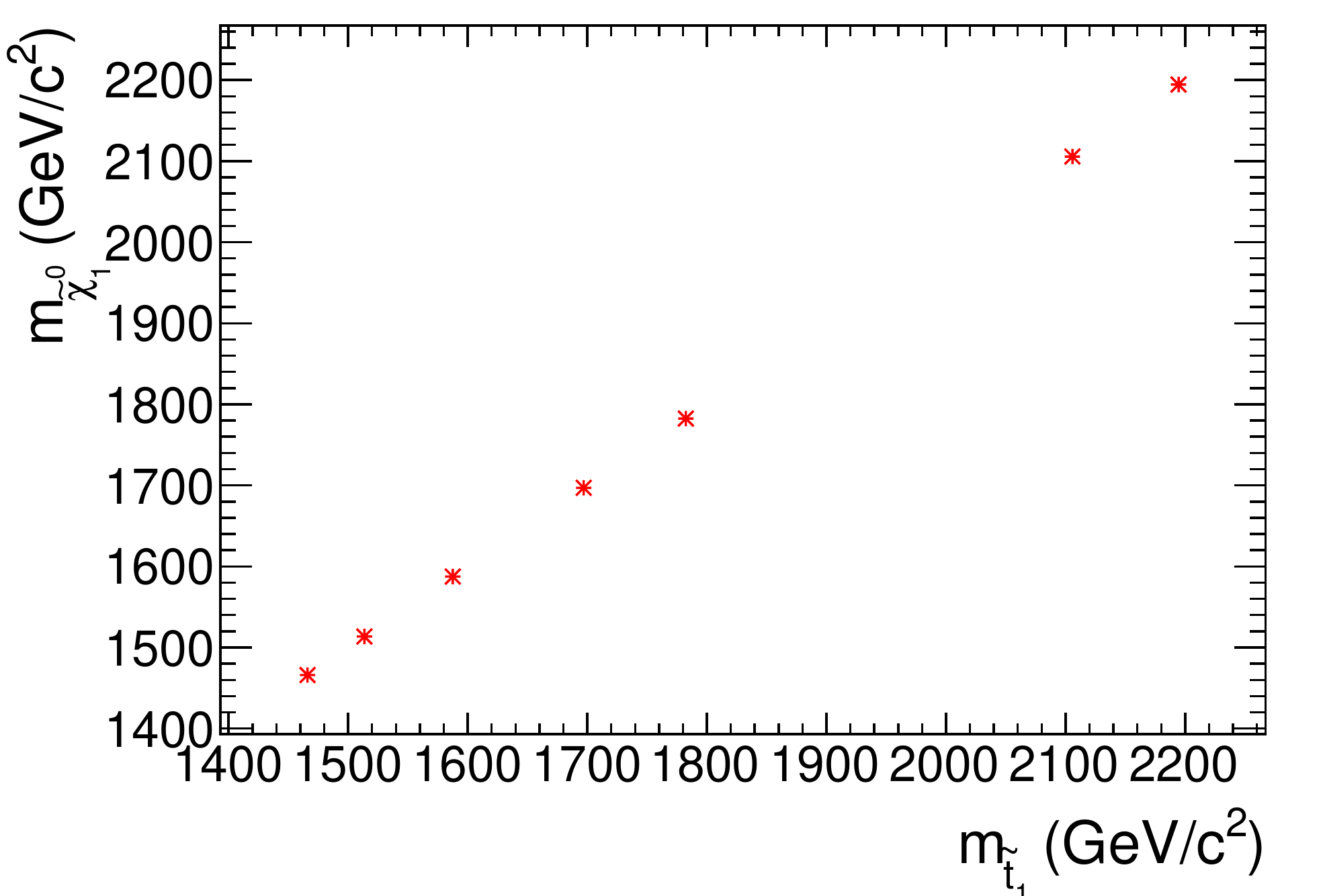}
}
\subfigure[$\tau_{\tilde{t}}$ vs $m_{\tilde{t}}-m_{\tilde{\chi}}$ for  $\tau_{\tilde{t}}>100$ ns, soft $\Omega_{\tilde{\chi}} $ cut.]{
\label{fig:sO_lt_stable}
\includegraphics[width=
0.45\textwidth, height= 4.7 cm]{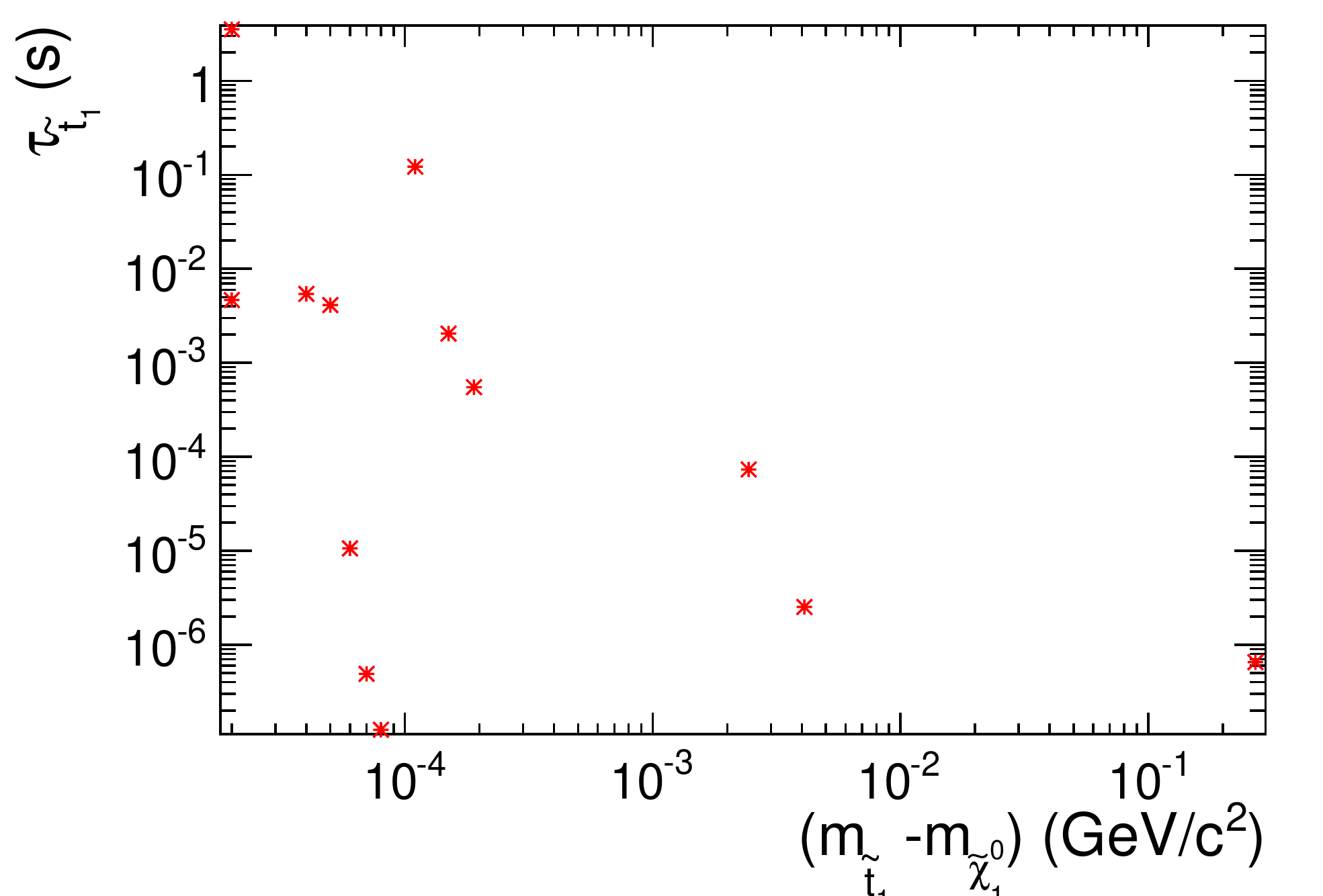}
}
\hskip 1 cm
\subfigure[$m_{\tilde{\chi}}$ vs $m_{\tilde{t}}$ for  $\tau_{\tilde{t}}>100$ ns, soft $\Omega_{\tilde{\chi}} $ cut.]{
\label{fig:sO_m_stable}
\includegraphics[width=
0.45\textwidth, height= 4.7 cm]{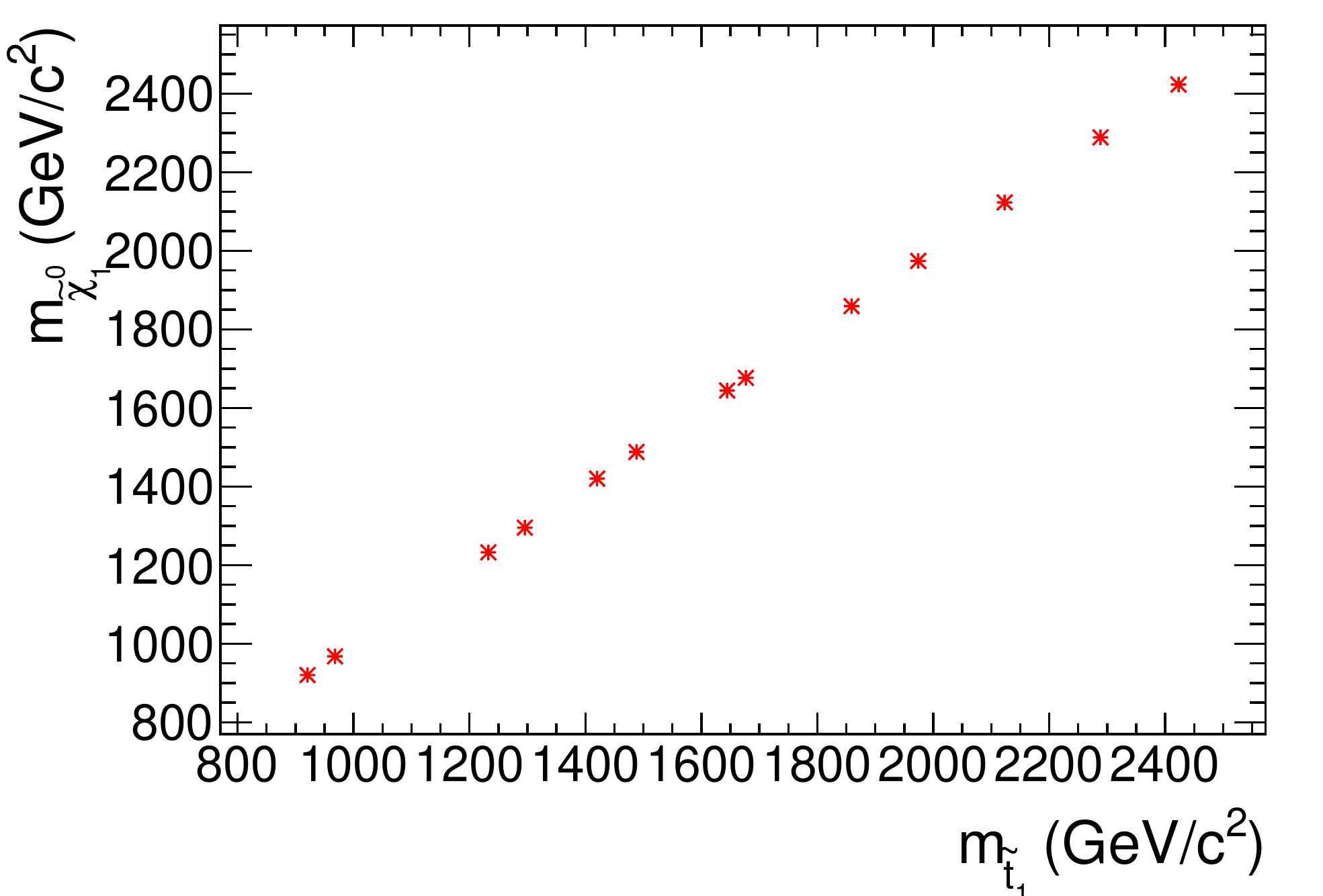}
}
\subfigure[$\tau_{\tilde{t}}$ vs $m_{\tilde{t}}-m_{\tilde{\chi}}$ for  $\tau_{\tilde{t}}>100$ ns, no requirement on $\Omega_{\tilde{\chi}} $.]{
\label{fig:nO_lt_stable}
\includegraphics[width=
0.45\textwidth, height= 4.7 cm]{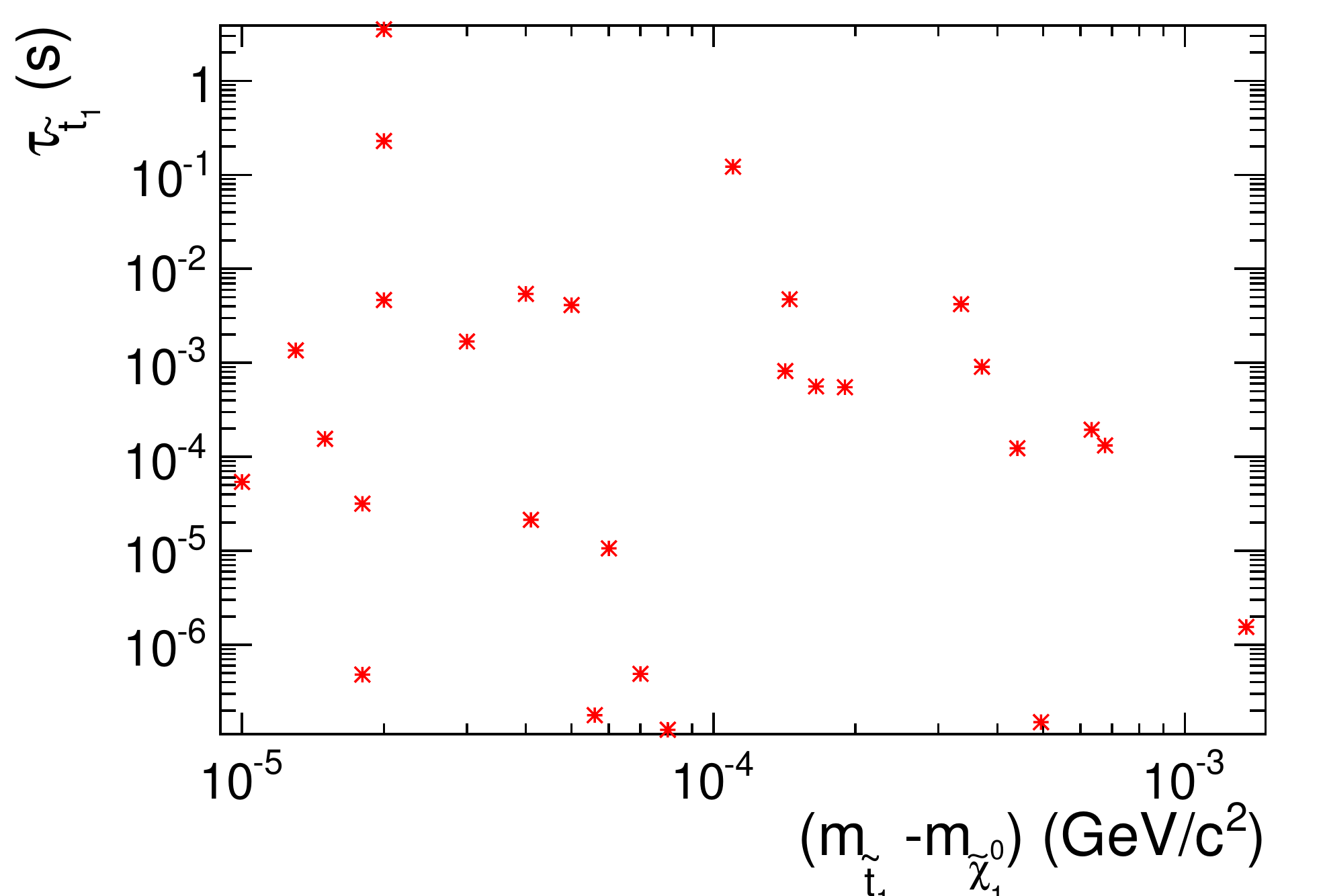}
}
\hskip 1 cm
\subfigure[$m_{\tilde{\chi}}$ vs $m_{\tilde{t}}$ for  $\tau_{\tilde{t}}>100$ ns, no requirement on $\Omega_{\tilde{\chi}}$.]{
\label{fig:nO_m_stable}
\includegraphics[width=
0.45\textwidth, height= 4.7 cm]{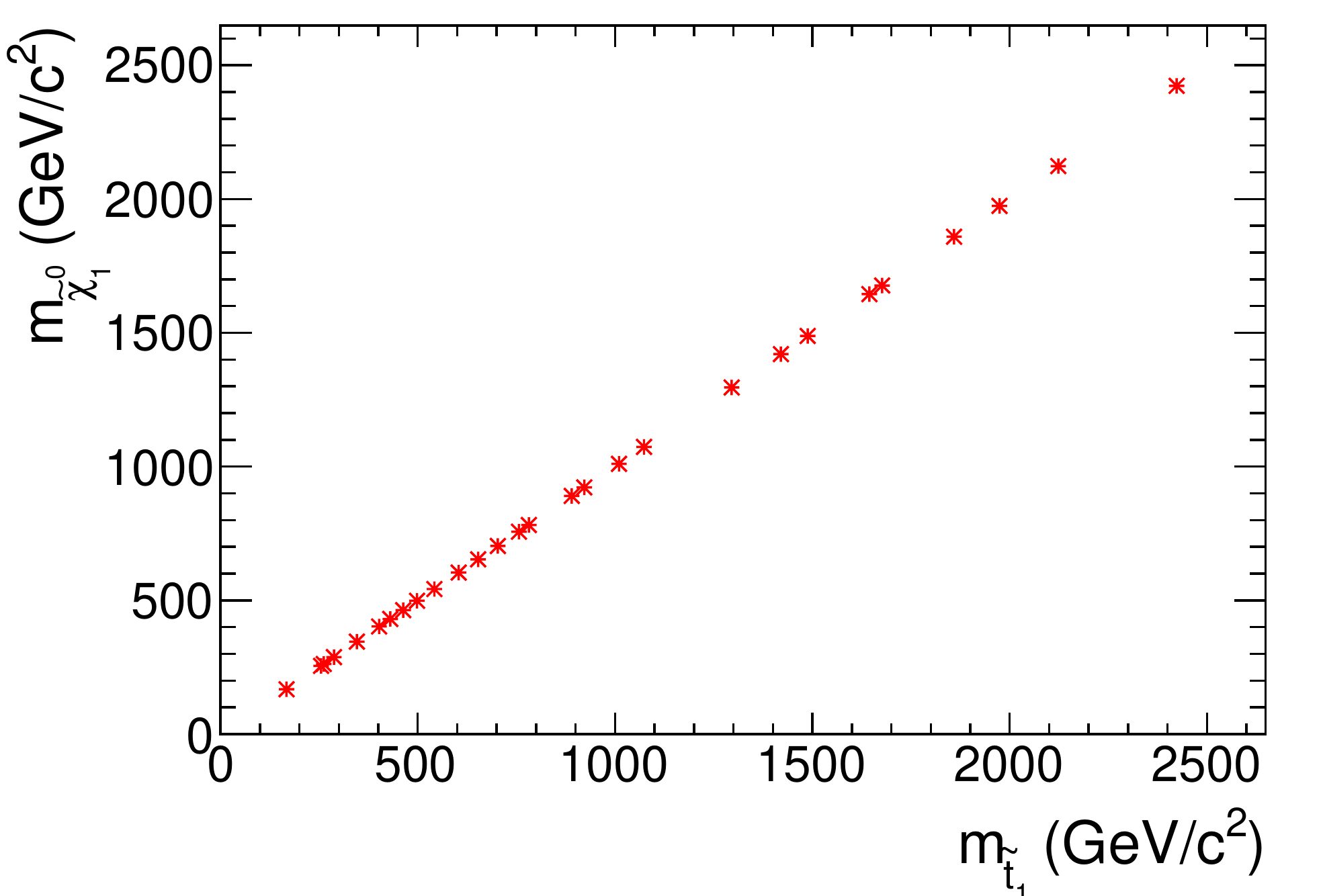}
}
\caption{\label{fig:lifetimeCut} The left column shows plots of the stop lifetime as a function of the mass difference between the stop and neutralino for the three different cuts on $\Omega_{\tilde{\chi}} h^{2}$. The right column shows plots of the neutralino mass versus the stop mass. All plots show models where the stop lifetime is greater than 100 ns.}
\end{figure}

\par Figure \ref{fig:lifetime} (left) shows the lifetime of the stop versus the stop-neutralino mass difference and (right)  the stop/neutralino mass difference versus the neutralino mass for the selected benchmark models. Figure \ref{fig:hO_lt} shows how the lifetime increases with decreasing stop-neutralino mass difference and also that for some models the mass difference is extremely small, at the order of $10^{-2}$ $\MeV$. The Figure also shows how the lifetimes range from around $10$ s to $10^{-26}$ s. We see that we can typically find models that pass all constraints for a variety of mass differences. Strong stop-neutralino
mass degenracy on the other hand can only be found for neutralino masses above 200 $\GeV$.

\par In more traditional SUSY models the stop is expected to be very short-lived with a lifetime around $10^{-24}$ s. From this perspective a majority of the models  in Figure \ref{fig:lifetime} give rise to long-lived stops. This study however focuses on stops which can both hadronise, traverse and leave a particle detector without decaying. The largest particle detector currently in use is ATLAS\cite{atlasCSC} which is 44 m long and has a diameter of 25 m. A particle moving with the speed of light will traverse the 22 metres from the interaction point to the end of ATLAS in 73 ns.  The stop masses in this study lies between 255 $\GeV$ to 2 $\TeV$ and can move with speeds less than the speed of light. A minimum stop lifetime is therefore chosen to be 100 ns, this lifetime  allows particles with a speed $v=0.7c$ to traverse the maximum flight path in ATLAS. From a detector analysis perspective this is quite conservative since most particles will move in the plane transverse to the beam-line giving a flight path much shorter than 22 m.
\par Figure \ref{fig:lifetimeCut} shows in the left column plots of the stop lifetime as a function of the stop-neutralino mass difference  and in the right column  plots of the neutralino mass as a function of the stop mass, after requiring the stop lifetime to be greater than 100 ns. The models are divided according to the hard (top), soft (middle) and no (bottom) $\Omega_{\tilde{\chi}} h^{2}$ cuts. The most striking tendency is how the number of models with a long-lived stop increases as the $\Omega_{\tilde{\chi}} h^{2}$ cut is made looser. Also note from the plots  \ref{fig:hO_m_stable} and \ref{fig:sO_m_stable} and  \ref{fig:nO_m_stable} how loosening the cut on $\Omega_{\tilde{\chi}} h^{2}$ allows for lighter stop-masses. This is understood by recalling that the neutralino-neutralino annihilation cross section increases with decreasing neutralino mass and that the neutralino-stop coannihilation cross section increases when the stop and neutralino are close in mass. When  respecting the relic density as measured by WMAP\cite{wmap} the contribution to the effective neutralino annihilation cross section from the neutralino-neutralino annihilation and the stop-neutralino coannihilation needs to  be balanced. For this reason only TeV-scale neutralinos are allowed to be close to mass degenerate with the stops when $\Omega_{\tilde{\chi}} h^{2}$ is restricted, as a comparison between Figures \ref{fig:hO_lt_stable} and  \ref{fig:hO_m_stable} as well as \ref{fig:sO_lt_stable} and  \ref{fig:sO_m_stable} show. Note that this is also the case when we allow the relic density to be more than $\pm 10\sigma$ away from the central WMAP value, corresponding to the soft $\Omega_{\tilde{\chi}} h^{2}$ requirement. When the relic density is unrestricted the contributions to the effective neutralino annihilation cross section also become unrestricted and low mass neutralinos can be close to degenerate with the stops.

\par As seen from each of  the plots \ref{fig:hO_lt_stable}, \ref{fig:sO_lt_stable} and \ref{fig:nO_lt_stable} the 100 ns cut leaves only models in which the neutralino and the stop are close to being mass-degenerate. The largest mass difference giving a long-lived stop is approximately 200 $\MeV$  and is found for the softer $\Omega_{\tilde{\chi}} h^{2}$ requirement in Figure \ref{fig:sO_lt_stable}. Some models with a mass difference at the order of  $\MeV$ are also found in \ref{fig:hO_lt_stable}. However, most models with a long lived stop have a stop-neutralino mass difference below 0.1 $\MeV$. The extreme cases are found in Figure \ref{fig:nO_lt_stable} where several models have a mass difference of the order $10^{-2}\, \MeV$.  Note that no models where the decay $\stop \rightarrow c\NO$ is open pass the 100 ns lifetime cut; all models passing this cut decay through $\stop \rightarrow u\NO$.

\par From this section it has become apparent that in order to get a long-lived
 stop in SUSY models with a neutralino LSP, the stop-neutralino mass
difference  has to be extremely small. Although it entails a certain
degree of fine-tuning, approximately mass-degenerate scenarios are
chosen. The motivation for this is that all non-excluded regions in the
parameter space need to be investigated and the chosen models would give
rise to distinctive experimental signatures.

\begin{figure}[tbp]
\centering
\subfigure[$\Omega_{\tilde{\chi}} h^{2}\! =\! 0.1099 \pm 2 \cdot 0.0062$.]{
\label{fig:hO_xsec}
\includegraphics[width=4.5 cm, height=4.5 cm]{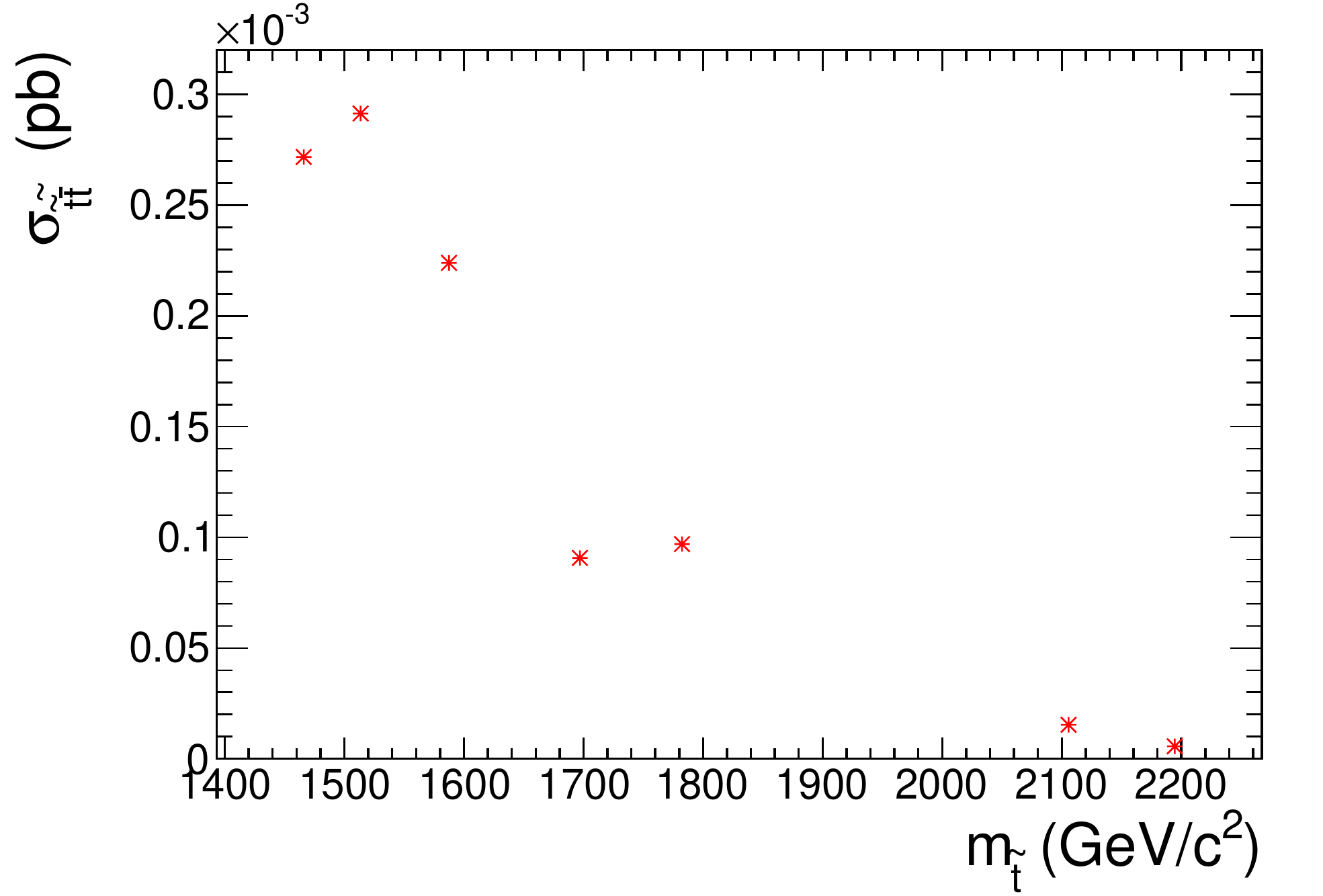}
}
\subfigure[$0.05<\Omega_{\tilde{\chi}} h^{2} < 0.2$.]{
\label{fig:sO_xsec}
\includegraphics[width=4.5 cm, height=4.5 cm]{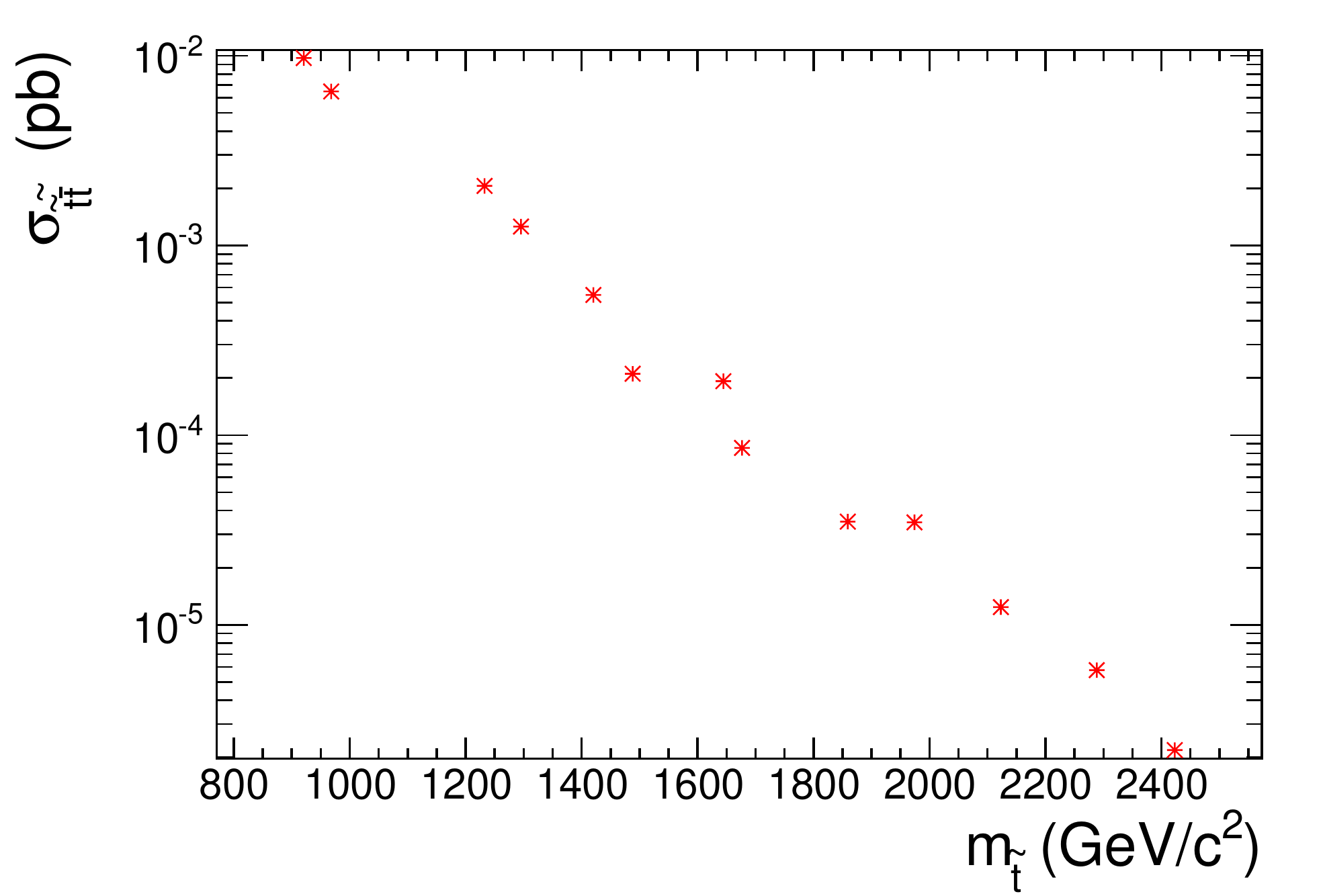}
}
\subfigure[No requirement on $\Omega_{\tilde{\chi}} h^{2}$. ]{
\label{fig:nO_xsec}
\includegraphics[width=4.5 cm, height=4.5 cm]{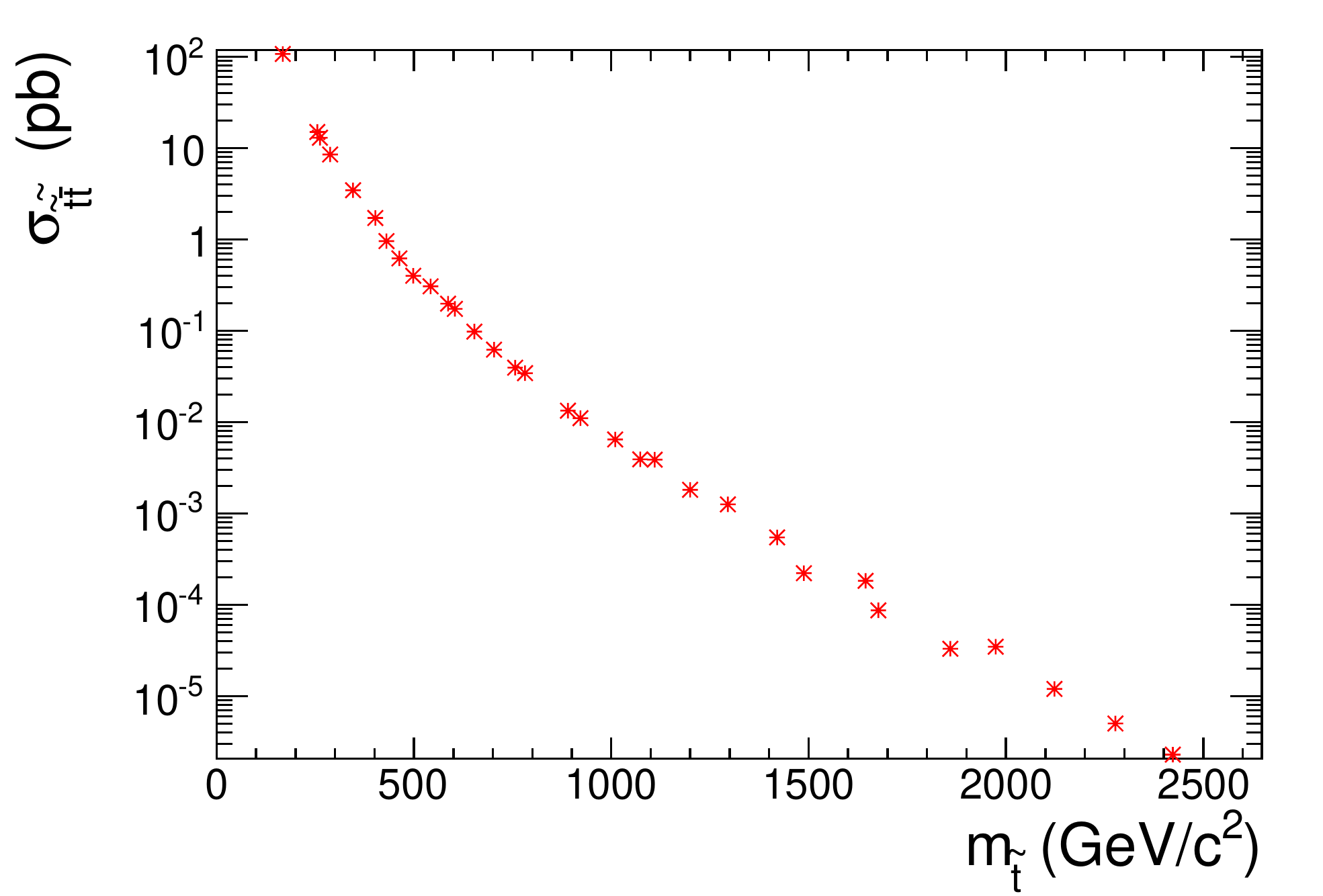}
}
\caption{\label{fig:xsec}Inclusive stop pair-production cross section as a function of the stop mass, for the three $\Omega_{\tilde{\chi}} h^{2}$ requirements.}
\end{figure}

\section{Phenomenology of long-lived stop models at the LHC}
Figure \ref{fig:xsec} shows the inclusive stop pair-production cross section for the models with a stop lifetime greater than 100 ns. The calculation was made with {\sc Pythia}. Relatively high cross sections of the order of $10$~pb for $pp$-collisions at $\sqrt{s}=14$ TeV can be achieved for low stop masses in the case where $\Omega_{\tilde{\chi}} h^{2}$ is unconstrained, see Figure \ref{fig:nO_xsec}. When $\Omega_{\tilde{\chi}} h^{2}$ is restricted the maximum cross sections are   $10^{-2}$  and $10^{-4}$ pb, see Figure \ref{fig:sO_xsec} and \ref{fig:hO_xsec} respectively. In all three plots the cross section falls rapidly with mass and for masses above 1 $\TeV$ it falls below 1 fb.
\par In models where the stop is the NLSP one would naively expect that  pair-production of stops would dominate the Beyond the Standard Model (BSM) physics processes at a particle collider experiment. However, a closer look at the main production mechanisms at a 14 TeV proton proton collider shows that this is not necessarily the case. The SUSY models can be categorised in three cases depending on which is the dominant BSM process;

\begin{figure}[tbp]
\centering
\subfigure[$f+ \bar{f}\rightarrow H^{+} + H^{-}$ ]{
\includegraphics[width=0.30\textwidth]{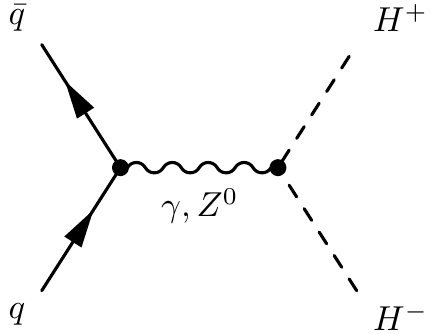}
}
\subfigure[$f+ \bar{f}'\rightarrow H^{\pm} + h^{0}, H^{0}$ ]{
\includegraphics[width=0.30\textwidth]{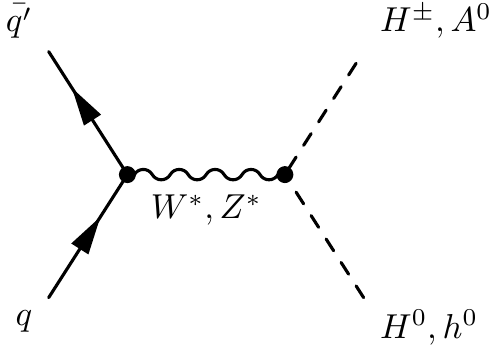}
}
\subfigure[$f+ \bar{f}\rightarrow A^{0} + h^{0}$ ]{
\includegraphics[width=0.30\textwidth]{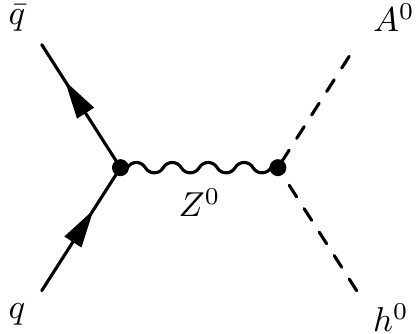}
}
\caption{\label{fig:higgsProd} Higgs pair-production mechanisms.}
\end{figure}

\begin{figure}[t]
\centering
\subfigure[ ]{
\label{fig:squarkgluino}
\includegraphics[width=0.25\textwidth]{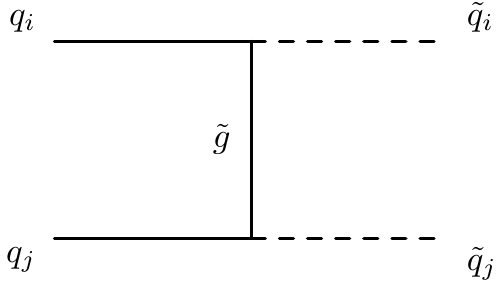}
}
\hskip 1 cm
\subfigure[ ]{
\label{fig:squarkneutralino}
\includegraphics[width=0.25\textwidth]{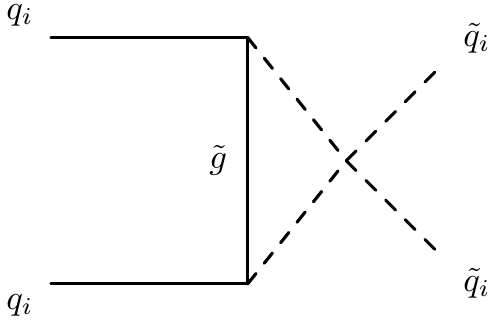}
}

\caption{\label{fig:squarkProd} Squark pair-production processes.}
\end{figure}

\begin{figure}[t]
\centering
\subfigure[ ]{
\label{fig:qaq1}
\includegraphics[width=0.25\textwidth]{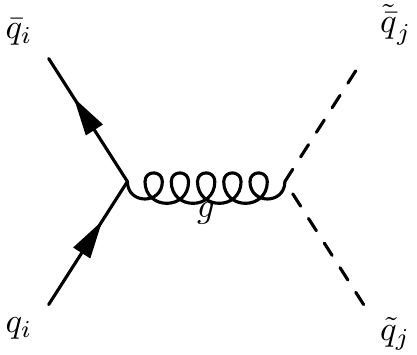}
}
\subfigure[ ]{
\label{fig:gluonsquark1}
\includegraphics[width=0.25\textwidth]{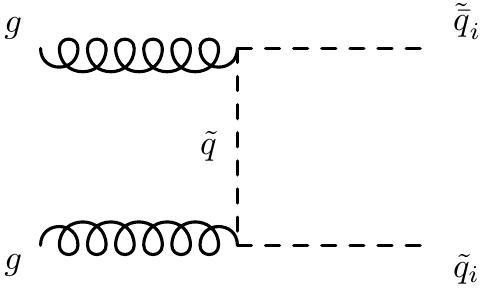}
}
\subfigure[ ]{
\label{fig:gluonsquark2}
\includegraphics[width=0.25\textwidth]{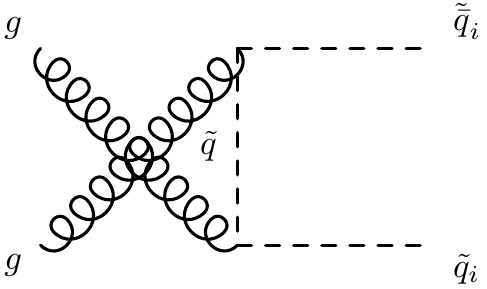}
}
\subfigure[ ]{
\label{fig:qaq2}
\includegraphics[width=0.25\textwidth]{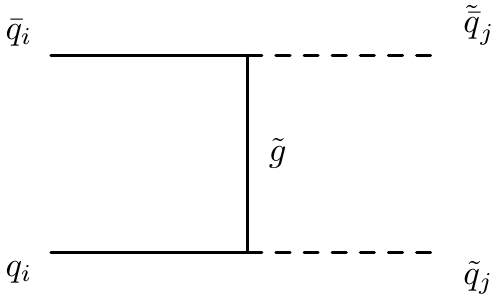}
}
\subfigure[ ]{
\label{fig:gluonsquark3}
\includegraphics[width=0.25\textwidth]{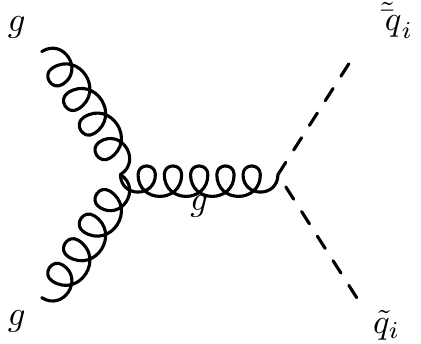}
}
\subfigure[ ]{
\label{fig:gluonsquark4}
\includegraphics[width=0.25\textwidth]{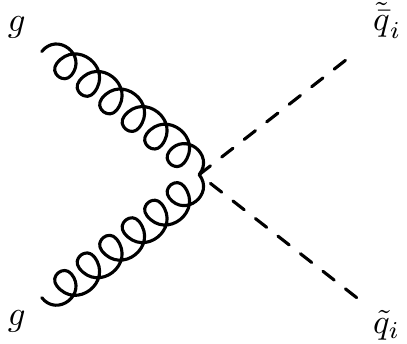}
}

\caption{\label{fig:qaq} Quark-anti-quark annihilation and gluon fusion diagrams producing a squark anti-squark pair.}
\end{figure}

\begin{itemize}
\item[1.]  Higgs production, mainly $f+\bar{f}^{'} \rightarrow H^{\pm} + h_{0} $ and $f+\bar{f} \rightarrow H^{\pm} + H_{0} $ accompanied by the production of the other Higgs particles, see Figure \ref{fig:higgsProd}.
\item[2.] First and second generation squark production, mainly $q_{i} + q_{j} \rightarrow \tilde{q}_{i,L\/R} + \tilde{q}_{j, L\/R}$, see Figure \ref{fig:squarkProd}.
\item[3.]  Third generation squark anti-squark production, mainly $g+g \rightarrow \tilde{t}_{1} + \tilde{\bar{t}}_{1}$, see Figure \ref{fig:qaq}
\end{itemize}

\noindent \textbf{Higgs-production:}  Figure \ref{fig:higgsProd} shows that the leading order Higgs
pair production mechanisms happens via exchange of $W$ or $Z$ bosons
in the $s$-channel.  The lighter neutral Higgs mass $m_{h^{0}}$ is
restricted from above and is expected be below 135 $\GeV$ in the MSSM \cite{mssm}.  The
heavier neutral Higgs $H^{0}$, the pseudoscalar $A^{0}$ and the
charged Higgs $H^{\pm}$ can be arbitrarily large.  However,
$m_{H^{0}}^{2}$ and $m_{H^{\pm}}^{2}$ can be expressed as functions of
$m_{A^{0}}^{2}$ and will be strongly dependent on the value set for
this parameter of the MSSM-8 model.  All the heavier Higgs masses will
in fact be very close to the value of $m_{A^{0}}$, usually not
separated by more than a few $\GeV$.

\noindent \textbf{Production of squark-pairs:} Figure \ref{fig:squarkProd} shows
the squark pair production mechanisms important in this work.
Figure \ref{fig:squarkProd} describes the strong interaction
processes and will in most cases dominate completely.  Squarks could also be produced
in electroweak processes through e.g. Higgs, gauge boson, neutralino or chargino exchange. However, the contribution from such processes is expected to be small compared to the strong production processes. We have verified that this is indeed the case for our models.

\noindent \textbf{Production of squark-antisquark pairs:} Squark anti-squark pairs
are produced via gluon-gluon fusion  shown in Figures \ref{fig:gluonsquark1},
\ref{fig:gluonsquark2}, \ref{fig:gluonsquark3} and
\ref{fig:gluonsquark4}.  Quark-anti-quark annihilation diagrams like
the ones shown in Figures \ref{fig:qaq1}  and \ref{fig:qaq2} also contribute. Even in this case, electroweak processes could contribute, but are expected to be subdominant compared to the strong processes. We have also here verified that this is the case for our models.

In practice, we calculate the production cross sections with {\sc Pythia}, including the strong processes given above \cite{Daw85}.

\par Whether a scenario is dominant depends on the masses of the
Higgs-particles, the squark sector and the gluino (recall that for
all these models the LSP and the lightest stop is very close in mass).
For models with relatively light squarks the strong processes in Figure
\ref{fig:qaq} dominate. Stop-antistop pair production will
dominate, since the first two generations squarks have considerably
higher mass.
For models where the squark masses are large, the gluon-fusion process
becomes suppressed by the low probability to obtain incoming gluons
with a large fraction of the proton momentum.
Instead, for intermediate values
of the squark masses, the strongly coupled process of Figure
\ref{fig:squarkProd} becomes leading. In this case the flavour of the
proton valence quarks translates into production of squark-pairs from
the first two generations.
For models where either the squark-masses or the mass
of the gluino-propagator gets very large, weak production of
Higgs-bosons, as in Figure \ref{fig:higgsProd}, will dominate.

\begin{table}[t] \footnotesize
\begin{tabular}{|c|c|c|c|c|c|ccc|}
\hline
\multirow{2}{*}{Case}&\multirow{2}{*}{Model} &$m_{\tilde{t}_{1} }$& $m_{\tilde{u}_{L} }$&$m_{A}$ & $m_{\tilde{g}}$ &\multicolumn{3}{|c|}{Dominating}  \\

& &  ($\GeV$) & ($\GeV$) & ($\GeV$)& ($\GeV$) &\multicolumn{3}{|c|}{processes} \\\hline
& & & & & & & & \\
& 1& 1514& 1752 & 557&17045 & & &  \\
\multirow{4}{*}{1} & 2 & 1697& 3287 &706  &11739&$f+ \bar{f}'$ &$\rightarrow$ &$H^{\pm} + h^{0}$ \\
& 3& 2194&3143 & 685 &15183 & $f+ \bar{f}' $ &$\rightarrow$ &$H^{\pm} + H^{0}$\\
&4  & 1466&2866& 595& 10146& $f+ \bar{f} $ &$\rightarrow$ &$A^{0} + h^{0}$\\
&5 &2106 &4923& 1281& 14571&$f+ \bar{f}$ &$\rightarrow$ &$H^{+} + H^{-}$ \\
&&& & & & &\\\hline
&&&& & & & \\
\multirow{4}{*}{2} & \multirow{4}{*} {6} &\multirow{4}{*} {1782} & \multirow{4}{*} {2173}& \multirow{4}{*} {1138}  & \multirow{4}{*} {12332}&$q_{i} + q_{j} $ &$\rightarrow$ &$ \tilde{q}_{i,R}+\tilde{q}_{j,R}$ \\
&  & & & &&$q_{i} + q_{j} $ &$\rightarrow$ &$ \tilde{q}_{i,L}+\tilde{q}_{j,L}$\\
&  & & & &&$q_{i} + \bar{q}_{j}$ &$\rightarrow$ &$\tilde{q}_{i,L}+\tilde{\bar{q}}_{j,R}$ \\
&  & & & & &$f+ \bar{f} $ &$\rightarrow$ &$\tilde{t}_{1} + \tilde{\bar{t}}_{1}$\\
&&&&& & & &\\\hline
& &&&& & & &\\
\multirow{4}{*}{3}&\multirow{4}{*}{7} &\multirow{4}{*} {1587} &\multirow{4}{*} {2795} & \multirow{4}{*} {1377} & \multirow{4}{*} {10986} &$g+g $ &$\rightarrow$ &$\tilde{t}_{1} + \tilde{\bar{t}}_{1}$\\
&  & & & & & $f+ \bar{f} $ &$\rightarrow$ &$\tilde{t}_{1} + \tilde{\bar{t}}_{1}$\\
&  & & & &&$g+g $ &$\rightarrow$ &$\tilde{b}_{1} + \tilde{\bar{b}}_{1}$\\
& & & & & &$g+g $ &$\rightarrow$ &$\tilde{b}_{2} + \tilde{\bar{b}}_{2}$\\
&&&&& & & &\\\hline
\end{tabular}
\caption{\label{tab:hO_prod}
Masses of the stop, sup, pseudoscalar Higgs, and gluino as well as the main production mechanisms for models with  $\Omega_{\tilde{\chi}} h^{2}=0.1099 \pm 2 \times 0.0062$.}
\end{table}

\begin{table}[ht] \footnotesize
\begin{tabular}{|c|c|c|c|c|c|ccc|}
\hline
\multirow{2}{*}{Case}&\multirow{2}{*}{Model} & $m_{\tilde{t}_{1} }$& $m_{\tilde{u}_{L} }$&$m_{A}$ & $m_{\tilde{g}}$&\multicolumn{3}{|c|}{Dominating} \\
& &  ($\GeV$) & ($\GeV$) & ($\GeV$)& ($\GeV$) &\multicolumn{3}{|c|}{processes}  \\\hline
 & & & & & & & & \\
& 1& 1488& 1571 & 245&10293& & & \\
\multirow{4}{*}{1} &2  & 1859& 3906 &852  &12859 &$f+ \bar{f}'$ &$\rightarrow$ &$H^{\pm} + h^{0}$\\
& 3& 2423&4451 & 1414 & 16766&$f+ \bar{f}' $ &$\rightarrow$ &$H^{\pm} + H^{0}$\\
& 4  & 2123&3996& 688& 14684&$f+ \bar{f} $ &$\rightarrow$ &$A^{0} + h^{0}$\\
& 5 &2288 &4259& 612&15830&$f+ \bar{f}$ &$\rightarrow$ &$H^{+} + H^{-}$ \\
&6&1420 & 2908& 569& 9822&&&\\
&&& & & & & &\\\hline
&&&& & & & & \\
\multirow{4}{*}{2}& 7 & 1644 & 1818& 816 & 11377 &$q_{i} + q_{j} $ &$\rightarrow$ &$ \tilde{q}_{i,R}+\tilde{q}_{j,R}$\\
& 8  &1974 & 2098 & 1301 &13656&$q_{i} + q_{j} $ &$\rightarrow$ &$ \tilde{q}_{i,L}+\tilde{q}_{j,L}$\\
& 9  &1295 &1355 &639 & 8962 &$q_{i} + \bar{q}_{j}$ &$\rightarrow$ &$\tilde{q}_{i,L}+\tilde{\bar{q}}_{j,R}$ \\
&10  &1677 &1846 &1229 & 11601&$g+ g $ &$\rightarrow$ &$\tilde{q}_{i,L} + \tilde{\bar{q}}_{i,L}$\\
& &&&& & & &\\\hline
& &&&& & & &\\
\multirow{4}{*}{3}& & &&  &  &$g+g $ &$\rightarrow$ &$\tilde{t}_{1} + \tilde{\bar{t}}_{1}$\\
& 11  &921 &4354 &432 & 12531&$f+ \bar{f} $ &$\rightarrow$ &$\tilde{t}_{1} + \tilde{\bar{t}}_{1}$\\
& 12 &1233 &4328 &623 & 8534 &$g+g $ &$\rightarrow$ &$\tilde{b}_{1} + \tilde{\bar{b}}_{1}$ \\
& 13 & 968 &3302 &587 &6702&$g+g $ &$\rightarrow$ &$\tilde{b}_{2} + \tilde{\bar{b}}_{2}$\\
& &&&& & & &\\\hline
\end{tabular}
\caption{\label{tab:sO_prod}
Masses of the stop, sup, pseudoscalar Higgs, and gluino as well as the main production mechanisms for models with  $0.05<\Omega_{\tilde{\chi}} h^{2} < 0.2$.}
\end{table}

\begin{table}[htbp] \footnotesize
\begin{tabular}{|c|c|c|c|c|c|ccc|}
\hline
\multirow{2}{*}{Case}&\multirow{2}{*}{Model} & $m_{\tilde{t}_{1} }$& $m_{\tilde{u}_{L} }$&$m_{A}$ & $m_{\tilde{g}}$ &\multicolumn{3}{|c|}{Dominating}\\
& &  ($\GeV$) & ($\GeV$) & ($\GeV$)& ($\GeV$) &\multicolumn{3}{|c|}{processes}\\\hline
& & & & & & & & \\
& 1& 1488&1571 & 245 &10293 & & & \\
\multirow{4}{*}{1}&2  &1859 &3906 &852 & 12859&$f+ \bar{f}'$ &$\rightarrow$ &$H^{\pm} + h^{0}$\\
& 3  &2423&4451&1414 & 16766&$f+ \bar{f}' $ &$\rightarrow$ &$H^{\pm} + H^{0}$\\
&4  &2123&3996&688& 14683&$f+ \bar{f} $ &$\rightarrow$ &$A^{0} + h^{0}$\\
& &&&& & & &\\\hline

&&&& & & & & \\
& 5  & 1644& 1818& 816&11377& & &\\
& 6 & 1974& 2098 & 1301&13654 & & &  \\
& 7 & 1295& 1355 & 639&8962 & & & \\
\multirow{4}{*}{2}& 8& 1677& 1846 &1229  &11601&$q_{i} + q_{j} $ &$\rightarrow$ &$ \tilde{q}_{i,R}+\tilde{q}_{j,R}$\\
& 9 &922&964&1398 &6378 &$q_{i} + q_{j} $ &$\rightarrow$ &$ \tilde{q}_{i,L}+\tilde{q}_{j,L}$\\
& 10  & 782&883& 983&5409 &$q_{i} + \bar{q}_{j}$ &$\rightarrow$ &$\tilde{q}_{i,L}+\tilde{\bar{q}}_{j,R}$\\
& 11 &703 &767& 1030&4865 & $g+ g $ &$\rightarrow$ &$\tilde{q}_{i,L} + \tilde{\bar{q}}_{i,L}$ \\
& 12 &890 & 1020&889 &6160& & & \\
& 13 &430 &464 &1183 &2976 & & & \\
& & &&& & & &\\\hline
& & &&& & & &\\

&14 & 346& 4754 &603 &2392 &&&\\
&15 & 542& 2523 &352 &3749&&&\\
&16 & 498& 2728 &1321 & 3447&&&\\
&17 & 463& 907 &1162& 3203&&&\\
\multirow{4}{*}{3}&18 &1073 &4483&1181  &7424&$g+g $ &$\rightarrow$ &$\tilde{t}_{1} + \tilde{\bar{t}}_{1}$  \\
&19  &604 &974&885 & 4176& $f+ \bar{f} $ &$\rightarrow$ &$\tilde{t}_{1} + \tilde{\bar{t}}_{1}$\\
&20 &756 &2885&1026 & 5231&$g+g $ &$\rightarrow$ &$\tilde{b}_{1} + \tilde{\bar{b}}_{1}$ \\
&21 & 1110&1528 &1107&6988&$g+g $ &$\rightarrow$ &$\tilde{b}_{2} + \tilde{\bar{b}}_{2}$\\
&22 & 653& 1242&363 &4516&&&\\
&23 & 255&4655 &1343 &1766&&&\\
&24 & 288&3030 &1268 &1993&&&\\
&25 & 402& 2910&1103 &2782&&&\\
& &&&& & & &\\\hline
\end{tabular}
\caption{\label{tab:nO_prod}
Masses of the stop, sup, pseudoscalar Higgs, and gluino as well as the main production mechanisms for models with no restrictions on  $\Omega_{\tilde{\chi}} h^{2}$.}
\end{table}

\par Table \ref{tab:hO_prod} lists the seven models with a long-lived stop and $\Omega_{\tilde{\chi}} h^{2}= 0.1099 \pm 2 \cdot 0.0062$. All models are given a bookkeeping number and the values of the stop\footnote{When referring to \emph{the} stop, we here mean the lightest stop, which we denote $\tilde{t}_{1}$.}, sup($\tilde{u}_L$), pseudoscalar Higgs  and gluino mass are listed. The models are divided into three cases according to which BSM production mechanism dominates. Case 1 represents Higgs production, case 2 first and second generation squark production and case 3 third generation squark production. For each main group of  production mechanisms the four largest contributing  processes are listed in the last column of the table.
\par As seen from Table \ref{tab:hO_prod} the BSM contribution for five of the models are dominated by Higgs production, one produces mainly squarks and one stop-anti-stop pairs. The five models which are dominated by Higgs production also have a large mass difference between the squark and the Higgs sector.  Also note that the gluino mass  for all five models is above 10 $\TeV$, which strongly suppresses the strong production of squarks that otherwise would dominate. This explains why the electroweak production of Higgs particles dominate the overall BSM particle production.
\par Table \ref{tab:sO_prod} shows the main production mechanisms for long-lived stop models with $0.05<\Omega_{\tilde{\chi}} h^{2} < 0.2$. Higgs production dominates six models, four are dominated by first and second generation squark production  and three models are dominated by third generation squark production. The underlying mechanisms determining which types of process dominate are the same as before.
\par When $\Omega_{\tilde{\chi}} h^{2}$ is unrestricted production of stop and sbottom dominates, closely followed by production of first and second generation squarks and a small fraction of models dominated by Higgs production, see Table \ref{tab:nO_prod}. The tendencies discussed before become even clearer for this choice of $\Omega_{\tilde{\chi}} h^{2}$.  Higgs and squark production dominate when the stop mass is quite high. Higgs production is prominent when $m_{A^0}$ is relatively small and squark production is suppressed due to heavy masses. When the Higgs masses increase and the squark masses decrease and the first and second generation squarks are comparable in mass to the third generation, the model is dominated by squark production. For models with smaller squark masses there can also  be stop-anti-stop production through gluon-gluon fusion. Third generation squark-anti-squark production dominates the BSM production  when the stop is much lighter than the other squarks and the heavy Higgs particles. However, as long as it is light enough for the gluon-gluon  fusion processes to be the most important production mechanism it will still dominate the overall BSM processes.
\par From Tables \ref{tab:hO_prod}-\ref{tab:nO_prod} we can deduce that models containing a long-lived stop with mass above 1500 $\GeV$ could be very hard to detect at the LHC. Not only because the cross section falls rapidly with mass, but also because these models are dominated by  Higgs production. In fact for some of these models it will be difficult to prove that supersymmetric processes are at work since none of the standard SUSY candles like high $p_{T}$ leptons or missing transverse energy will be present. The Higgs particles produced will decay to $b\bar{b}$ or $t\bar{t}$ and if the first decay dominates it will be hard to distinguish from background. For some of these models it could also happen that a long-lived stop signature will be the first indication of the presence of beyond the Standard Model physics.

\par Models dominated by first and second generation squark production can also be quite challenging since the squarks will rapidly decay through the mode $\tilde{q}_{i} \rightarrow \tilde{\chi}_{1}^{0} + q_{i}$ . These decays will give rise to hard jets and missing transverse energy. This signature will surely deviate from standard model processes, but due to the high squark masses in these models discovery  will require 14 TeV collisions and high statistics. In most of these models approximately 6\% of the total BSM processes produced in $pp$ collisions are stop-pair-production.  For this reason and of course depending on the mass and the cross section there should be a possibility of discovering such states.
\par Long-lived stops are of course easiest to discover in models where stop production dominates the overall SUSY processes. The limitations when it comes to discovering such states  arise from the cross section and how well the signal signature can be separated from the Standard Model background. This will be studied in more detail in the following sections.

\section{Discovering long-lived stops at the LHC}
If long-lived stops exist it will be of great importance to discover them at the LHC. In the following sections we will use the performance of the ATLAS\cite{atlasCSC} detector as  an example of a general purpose detector and investigate the possibility of discovering long-lived stops using generic and global cut variables. All events are generated using {\sc Pythia} 6.4.
\subsection{Interactions of long-lived stops}
Long-lived stops produced at a hadron collider will immediately form $R$-hadrons.  Both $R$-mesons $\tilde{t}_{1}\bar{q}$ and R-baryons $\tilde{t}_{1}qq$ can be formed.  We adapt the assumption that 90\% of the stops will initially form $R$-mesons and  10\% form $R$-baryons from Ref. \cite{kraan}. Through interactions with the detector material most mesons convert to baryons. In a conservative assumption any of the baryon states $\tilde{t}_{1}uu$ (charge$=+2e$), $\tilde{t}_{1}ud$ ($+e$) or $\tilde{t}_{1}dd$ ($0e$) are equally likely for a stop exiting the detector. The situation for the anti-stop is somewhat different; it will also form an $R$-meson when produced, but a transition to a anti-baryon state like $\tilde{\bar{t}}_{1}\bar{u}\bar{u}$ is unlikely and if it occurs it will lead to an immediate annihilation back to a meson. An anti-stop exiting the detector will therefore be in a mesonic  $\tilde{\bar{t}}_{1}u$ or  $\tilde{\bar{t}}_{1}d$ state, with charges $0$ and $-e$, respectively. Given these assumptions, roughly $80$\% of stop events are expected to have at least one charged $R$-hadron leaving the detector.
\par Due to ionisation energy losses, the charged $R$-hadrons will leave tracks in the detector's tracking devices. Very little energy will be deposited in the calorimeters because of an expected small interaction cross section\cite{kraan,deBoer:2007ii,Mackeprang:2006gx,Mackeprang:2009ad}. The interactions that do take place here are responsible for the meson to baryon conversion which can also lead to a change of electrically neutral states to charged states and vice versa. Finally, charged  $R$-hadron states also leave tracks in the outermost parts of a detector which are designed to detect muons.
\par An $R$-hadron remaining charged when traversing  a detector will have a similar signature as a high transverse momentum ($p_{T}$), possibly slow moving, muon. Jet activity in the event is expected to originate from  initial/final state radiation or from stop production through decay of heavier sparticles. Since there is a probability for the $R$-hadron to acquire or lose its charge, there can also be events where a high $p_{T}$ inner detector track disappears before reaching the muon detectors, or that a  high $p_{T}$ muon track has no inner detector track matching it. These effects are best studied with  more detailed detector simulations as in Ref. \cite{atlasCSC}, and are outside the scope of this paper.
\begin{table}[t]\footnotesize
\centering
\begin{tabular}{|c|c|c|c|c|} \hline

 \multirow{2}{*}{Model}& $m_{\tilde{t}_{1}}$&$\sigma_{\tilde{t}_{1}\tilde{\bar{t}}_{1}}$ &\multirow{2}{*}{N(100 fb$^{-1}$)} &\multirow{2}{*}{$\Omega_{\tilde{\chi}} h^{2}$}\\
 & $\GeV$ & (fb)& & \\\hline
1 & 255  &15000& 1.5$\cdot 10^{6}$ & 0.0019\\
2 & 542  &306.6& 30660 & 0.0047\\
3 & 746  &39& 3900 &0.0128\\
4 & 921  &9.7& 973& 0.0614\\
5  &1073&  3.9&390 & 0.0442\\
6  & 1588& 0.22& 22 & 0.1134\\\hline
\end{tabular}
\caption{\label{tab:stopProd} Inclusive stop pair cross section, expected number of events at 100 fb$^{-1}$ and $\Omega_{\tilde{chi}} h^{2}$ for the six selected SUSY models.}
\end{table}

\subsection{Selection of benchmark models}
 A set of six models with stop masses ranging from 255 to 1588 $\GeV$ was selected.  The mass spectra and some details of these models are listed in Tables \ref{tab:stopProd} and  \ref{tab:stopPairProd} respectively.

The models were selected to provide a continuous and wide mass range.  All samples are scaled to 100 fb$^{-1}$ corresponding to one year of running 14 TeV collisions at a luminosity of $10^{-34} \textrm{cm}^{-2}\textrm{s}^{-1}$. Some models, for example those containing the 255 and 542 $\GeV$ stops, have, in principle, large enough cross sections to make earlier discoveries possible, although this has not  been investigated in this work.
\par All selected models have third generation squark production as the main BSM production mechanism. The total fraction of stops produced in a given model depends not only on the direct production of stop pairs but also on the indirect production via decay of heavier sparticles. The largest contributors to the latter are the production and decays of the other third generation squarks. These are produced via the processes $g + g \rightarrow \tilde{Q}_{i} + \tilde{\bar{Q}}_{i}$ and $f + \bar{f} \rightarrow  \tilde{Q}_{i} + \tilde{\bar{Q}}_{i}$, where $Q_{i}$ denotes $\tilde{t}_{1}$, $\tilde{t}_{2}$, $\tilde{b}_{1}$ or $\tilde{b}_{2}$. Table \ref{tab:stopPairProd} shows the contribution from each of the third generation squark production processes  to the total BSM cross section. The fractional numbers represent, in the cases of  $\tilde{t}_{2}$ and $\tilde{b}_{1,2}$,  the decays to a $\tilde{t}_{1}\tilde{\bar{t}}_{1}$ pair. For example, in model 2   $\tilde{b}_{1}\tilde{\bar{b}}_{1}$ production will constitute 13\% of the total BSM cross section and 96\% of all $\tilde{b}_{1}\tilde{\bar{b}}_{1}$ pairs will decay to $\tilde{t}_{1}\tilde{\bar{t}}_{1}$. Based on these numbers one can calculate the inclusive $\tilde{t}_{1}\tilde{\bar{t}}_{1}$ cross section ($\sigma_{\tilde{t}_{1}\tilde{\bar{t}}_{1}}$) and then get  the expected number of $\tilde{t}_{1}\tilde{\bar{t}}_{1}$ pairs at 100 fb$^{-1}$ (N(100 fb$^{-1})$) as shown in Table \ref{tab:stopProd}. The last column of Table \ref{tab:stopProd} list the value of $\Omega_{\tilde{\chi}} h^{2}$ for each model, which in all cases except for model 6 is well below the limit computed from the WMAP-measurement.
\begin{table}[t]\footnotesize
\centering
\begin{tabular}{|c|c|c|c|c|c|c|c|c|c|}
\hline
\multirow{2}{*}{Model}& $m_{\tilde{t}_{1}}$&  \multicolumn{8}{c|}{Percentage of total / Fraction decaying to $\tilde{t}_{1}\tilde{\bar{t}}_{1}$} \\\cline{3-10}
& $\GeV$&\multicolumn{2}{c|}{$\tilde{t}_{1}\tilde{\bar{t}}_{1}$}  & \multicolumn{2}{c|}{$\tilde{t}_{2}\tilde{\bar{t}}_{2}$} & \multicolumn{2}{c|}{$\tilde{b}_{1}\tilde{\bar{b}}_{1}$} & \multicolumn{2}{c|}{$\tilde{b}_{2}\tilde{\bar{b}}_{2}$} \\\hline
1 & 255 &92\% & 1 & 0.3\% & 0.36&  6 \% & 0.48 & 1 \% & 0.77 \\
2 & 542 &72\% & 1 & 2\% &0.36  &  13\% & 0.96 & 10\% &0.98 \\
3 &746 &39\% & 1 & 11\%& 0.16&  35\% & 0.00 & 14\%& 0.74 \\
4 &921 &23\% & 1 & 15\%& 0.00&  23\%& 0.32  & 20\%& 0.00 \\
5 &1073&38\% & 1 & 12\% & 0.19& 34\%& 0.00 & 15\%&0.81\\
6 &1588&30\% &1 & 11\%&0.62 &  20\% & 0.00 & 18\%&0.92 \\\hline
\end{tabular}
\caption{\label{tab:stopPairProd} The percentage of third generation squark production contributing to the total BSM cross section combined with  the fraction of stop and sbottom events decaying to $\tilde{t}_{1}\tilde{\bar{t}}_{1}$.}
\end{table}

\begin{table} \footnotesize
\centering
\begin{minipage}[b]{0.45\linewidth}\centering
\begin{tabular}{|c|c|c|}
\hline
\multirow{2}{*}{Process} & $\sigma$ & N(100 fb$^{-1}$) \\
 & (pb) & $10^{6} $\\\hline
 $Z$+ jet & 138.4 & 14 \\
 $W$+jet & 1192 & 119 \\
 $WW$, $ZW$, $ZZ$ & 231 & 23 \\
 $t\bar{t}$ & 490 & 49 \\
  $b\bar{b}$ & 5063 & 506 \\
  QCD J5 & 12490 & 1249 \\
  QCD J6 & 359 & 36 \\
  QCD J7 & 16 & 2 \\
  QCD J8 & 0.024 & 0.0024 \\\hline
\end{tabular}
\caption{\label{tab:bg}Background cross sections and expected number of events at 100 fb$^{-1}$.}
\end{minipage}
\hskip 1 cm
\begin{minipage}[b]{0.45\linewidth}
\centering
\begin{tabular}{|c|r|}
\hline
\multirow{2}{*}{Label} & $p_{T}$ interval \\
& ($\GeVm$) \\\hline
QCD J4 & 140-180 \\
QCD J5 &  280-560\\
QCD J6 & 560-1120 \\
QCD J7 & 1120-2240 \\
QCD J8 & $> 2240$ \\\hline
\end{tabular}
\caption{\label{tab:qcd} $p_{T}$-range of the hard scattering process defining the QCD kinematic slices.}
\end{minipage}
\end{table}

\subsection{Standard Model background processes}
All Standard Model processes giving rise to high-$p_{T}$ muons could mimic the signal of a long-lived, hadronising stop. The most important processes are:
\begin{itemize}
\item[a)] Z + jet
\item[b)] W + jet
\item[c)] $WW$, $ZW$, $ZZ$ (diboson production)
\item[d)] $t\bar{t}$
\item[e)]  $b\bar{b}$
\item[f)]  high $p_{T}$ QCD-processes not covered in the above list.
\end{itemize}

Muons are produced in the processes $Z \rightarrow \mu^{+}\mu^{-}$ and $W^{\pm} \rightarrow \mu^{\pm} \nu_{\mu} (\bar{\nu}_{\mu})$. In the $2 \rightarrow 1$ processes producing single vector bosons, the muons typically carry a momentum which is half of the boson mass. These muons do not have high enough $p_{T}$ to constitute any real background for the long-lived stops. In the $2 \rightarrow 2$ processes where the boson is recoiling against a jet, the boson itself carry momentum which is transferred to its decay products. This allows for  high-$p_{T}$ muons coming from $Z$ and $W$ decay and is the reason for the choice of a) and b).
\par The top quark decays to a $W$ and $b$-quark and gives rise to muons either through the decay of the $W$  or the decay of a $B$-meson, or both. In bottom pair-production all muons arise from the decay of $B$-mesons.
\par Background f) is referred to as high-$p_{T}$ QCD-processes. These processes produce the light quark flavours and/or gluons giving rise to multiple jets. Muons arise primarily from decays of charged pions and kaons. These processes  will be dominating  at the LHC and in order to  generate the number of events needed to ensure reasonably small statistical uncertainties it is common to divide events into $p_{T}$ intervals according to the hard scattering. Here we use the same division as in Ref. \cite{atlasCSC} and the intervals are  listed in Table \ref{tab:qcd}. Only the samples J5-J8 will be used in the following analysis as it turns out that the J4-background is negligible.
%%%%%%%%%%%%%%%%%%%%%%%%%%%%%%%%%%%%%%%%%%%%%%%%%%
%****************************DATA SELECTION/ACCEPTANCE CUTS********************************%
%%%%%%%%%%%%%%%%%%%%%%%%%%%%%%%%%%%%%%%%%%%%%%%%%%
\begin{figure}[htbp]
\subfigure[$p_{T}$-distribution for $R$-hadrons.]{\label{fig:pt_signal}
\includegraphics[width= 0.40\textwidth]{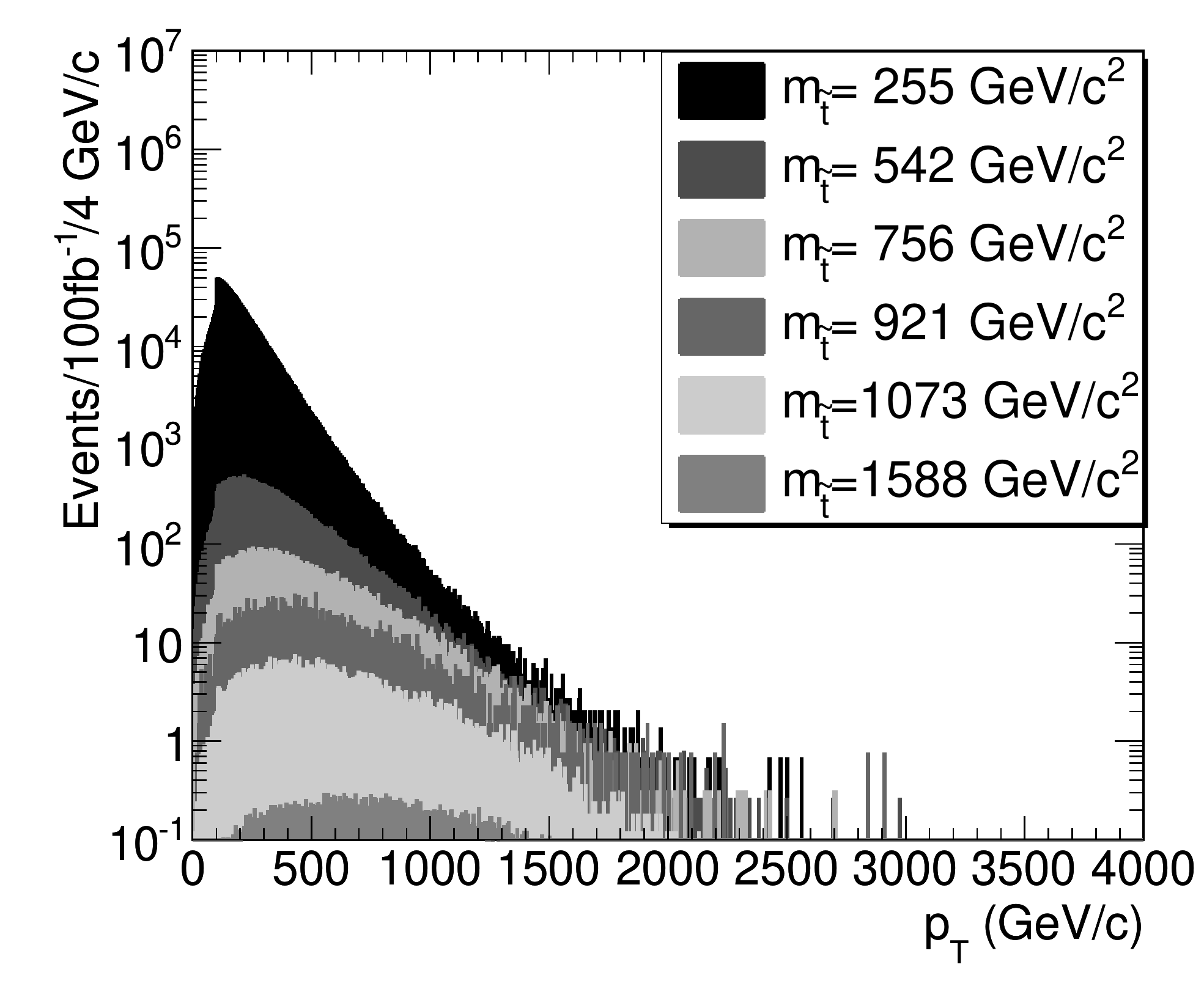} }
\hskip 1 cm
\subfigure[$p_{T}$-distribution for muons.]{\label{fig:pt_bg}
\includegraphics[width= 0.40\textwidth]{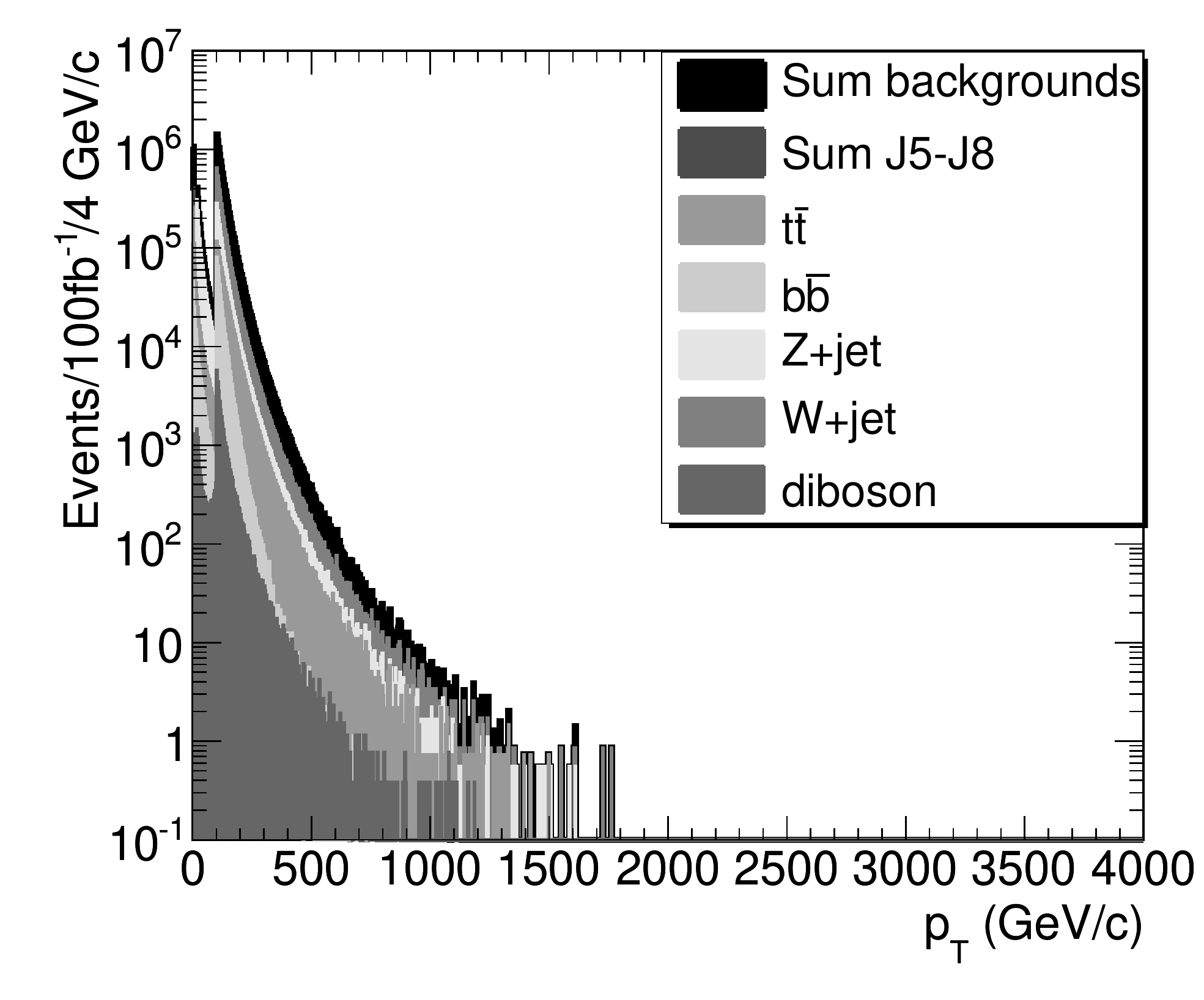} }

\subfigure[$\eta$-distribution for $R$-hadrons.]{\label{fig:eta_signal}
\includegraphics[width= 0.40\textwidth]{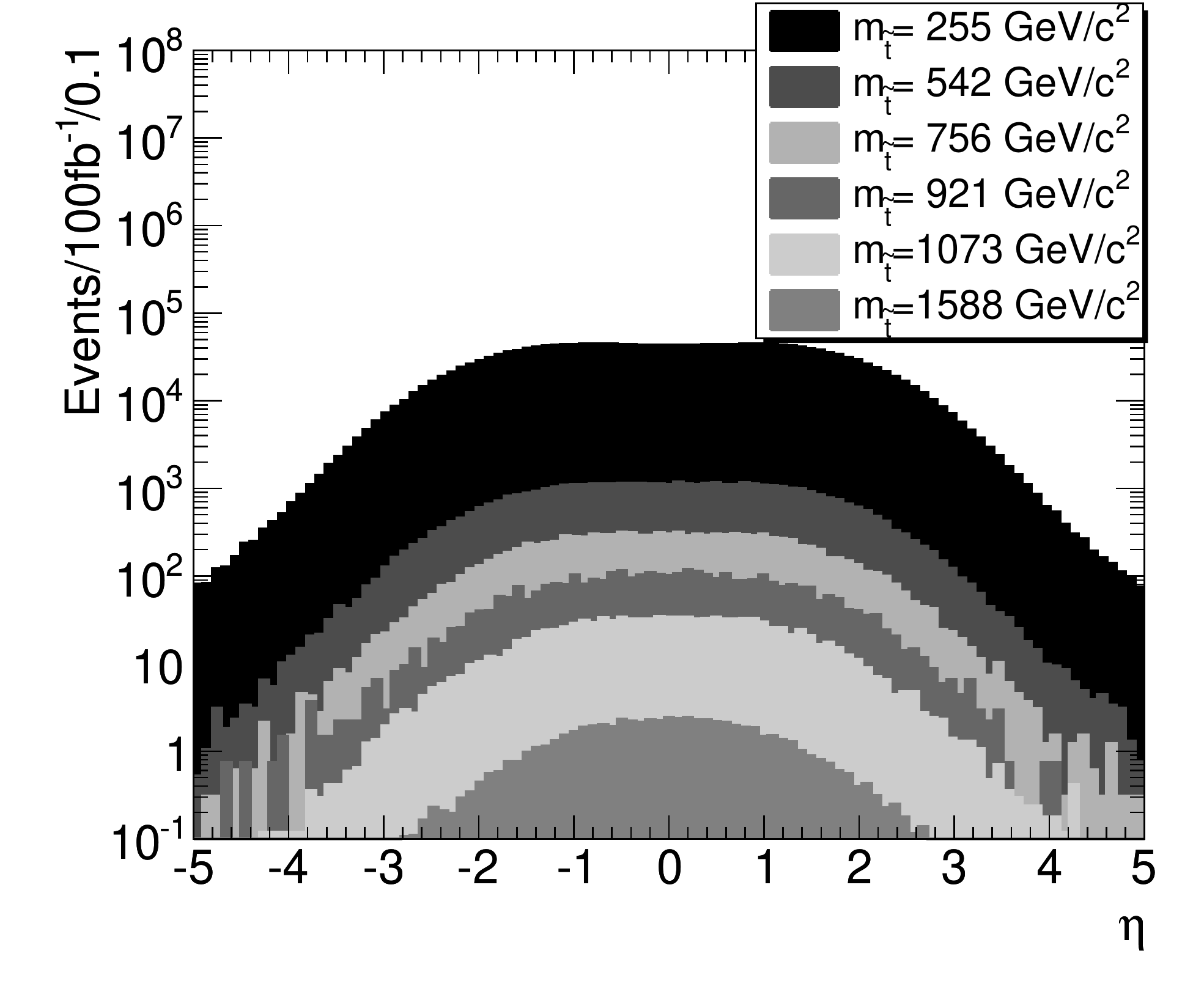} }
\hskip 1 cm
\subfigure[$\eta$-distribution for muons.]{\label{fig:eta_bg}
\includegraphics[width= 0.40\textwidth]{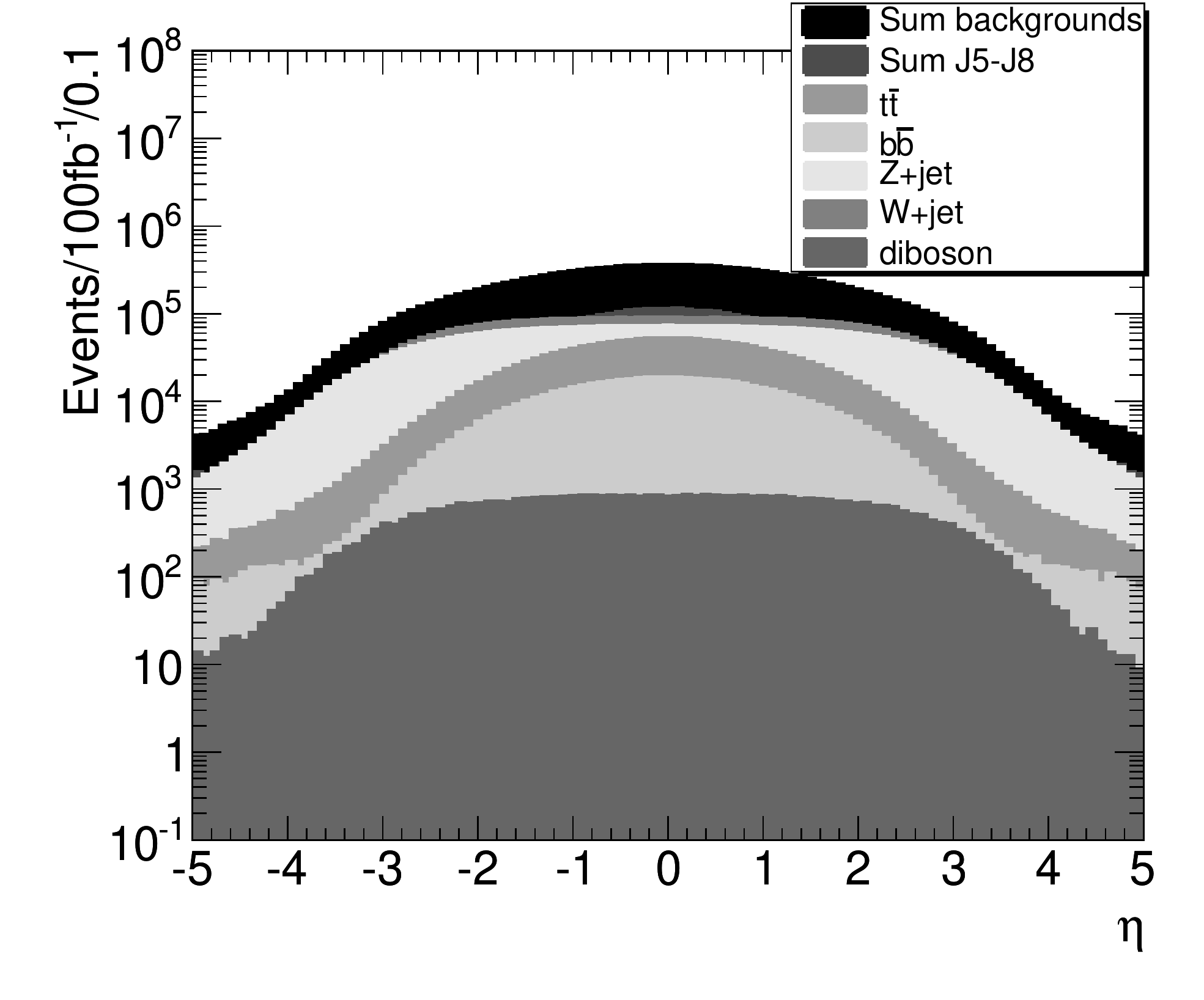} }

\subfigure[$\beta$-distribution for $R$-hadrons.]{\label{fig:beta_signal}
\includegraphics[width= 0.40\textwidth]{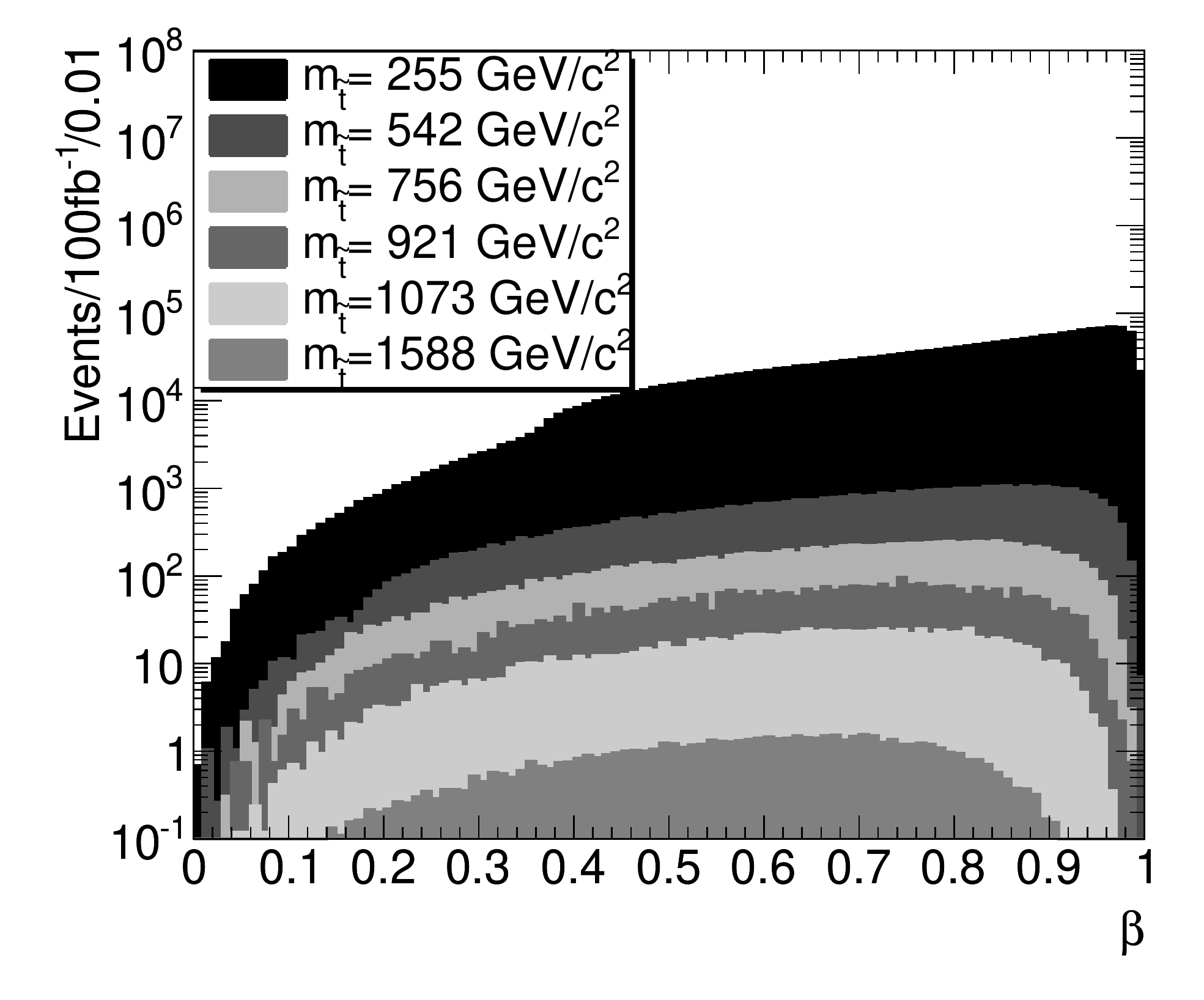}}
\hskip 1 cm
\subfigure[$\beta$-distribution for muons.]{\label{fig:beta_bg}
\includegraphics[width= 0.40\textwidth]{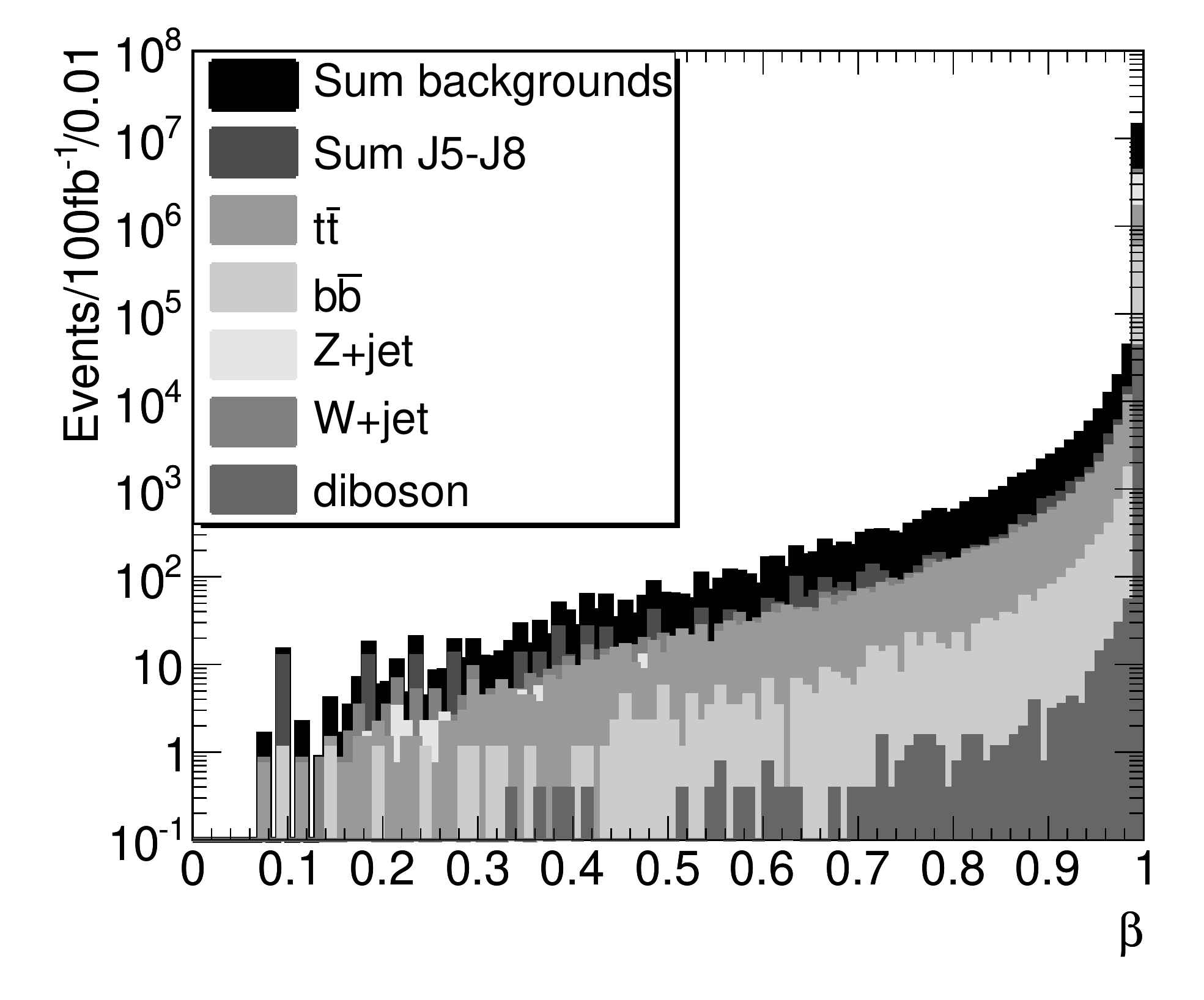} }
\caption{\label{fig:acceptance} The transverse momentum, $\eta$ and $\beta$ distributions for signal $R$-hadrons \protect \ref{fig:pt_signal},  \protect \ref{fig:eta_signal},  \protect \ref{fig:beta_signal} and the various muon backgrounds  \protect \ref{fig:pt_bg},  \protect \ref{fig:eta_bg},  \protect \ref{fig:beta_bg}.  All plots show the spectrum after a preselection including only events with at least one $R$-hadron/muon that has a transverse momentum larger than 100 $\GeVm$. }
\end{figure}

\begin{figure}
\subfigure[Summed $p_{T}$ of tracks in a cone of size  $\Delta R<0.2$ around the $R$-hadrons.]{\label{fig:ptcone_signal}
\includegraphics[width= 0.40\textwidth]{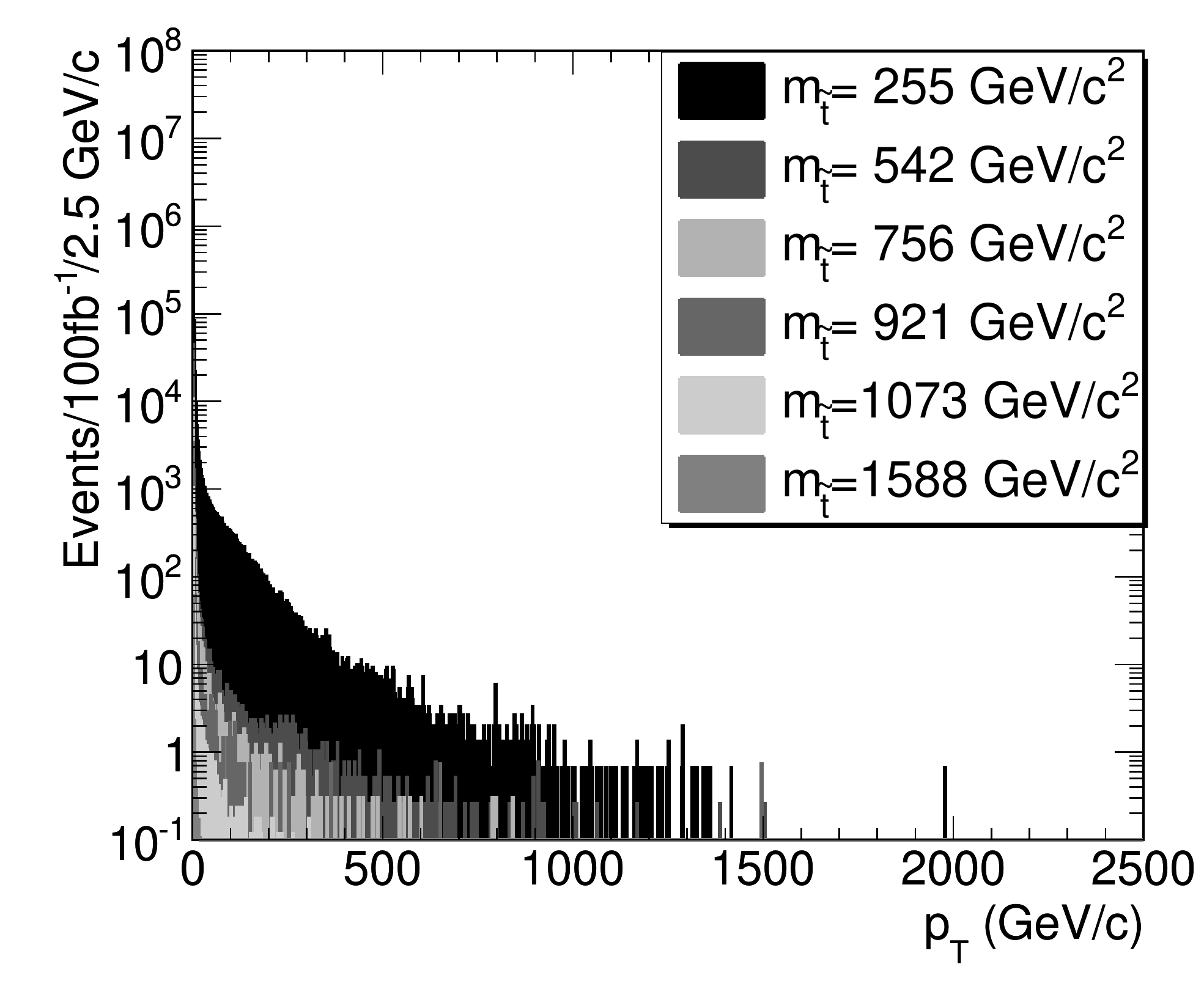}}
\hskip 1 cm
\subfigure[Summed $p_{T}$ of tracks in a cone of size $\Delta R<0.2$ around the muons.]{\label{fig:ptcone_bg}
\includegraphics[width= 0.40\textwidth]{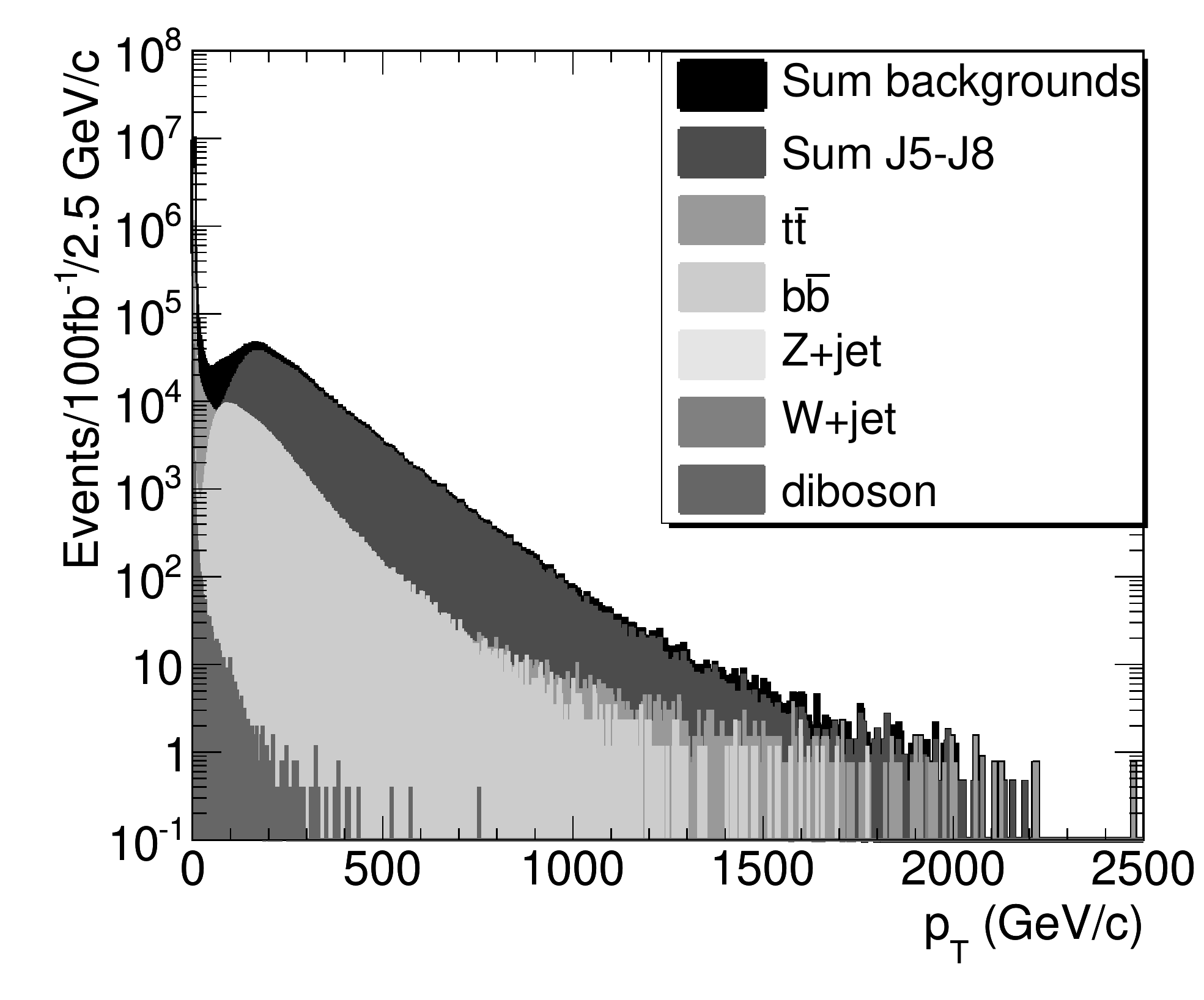} }

\subfigure[Distance $\Delta R$ between $R$-hadron and closest jet.]{\label{fig:closestjet_signal}
\includegraphics[width= 0.40\textwidth]{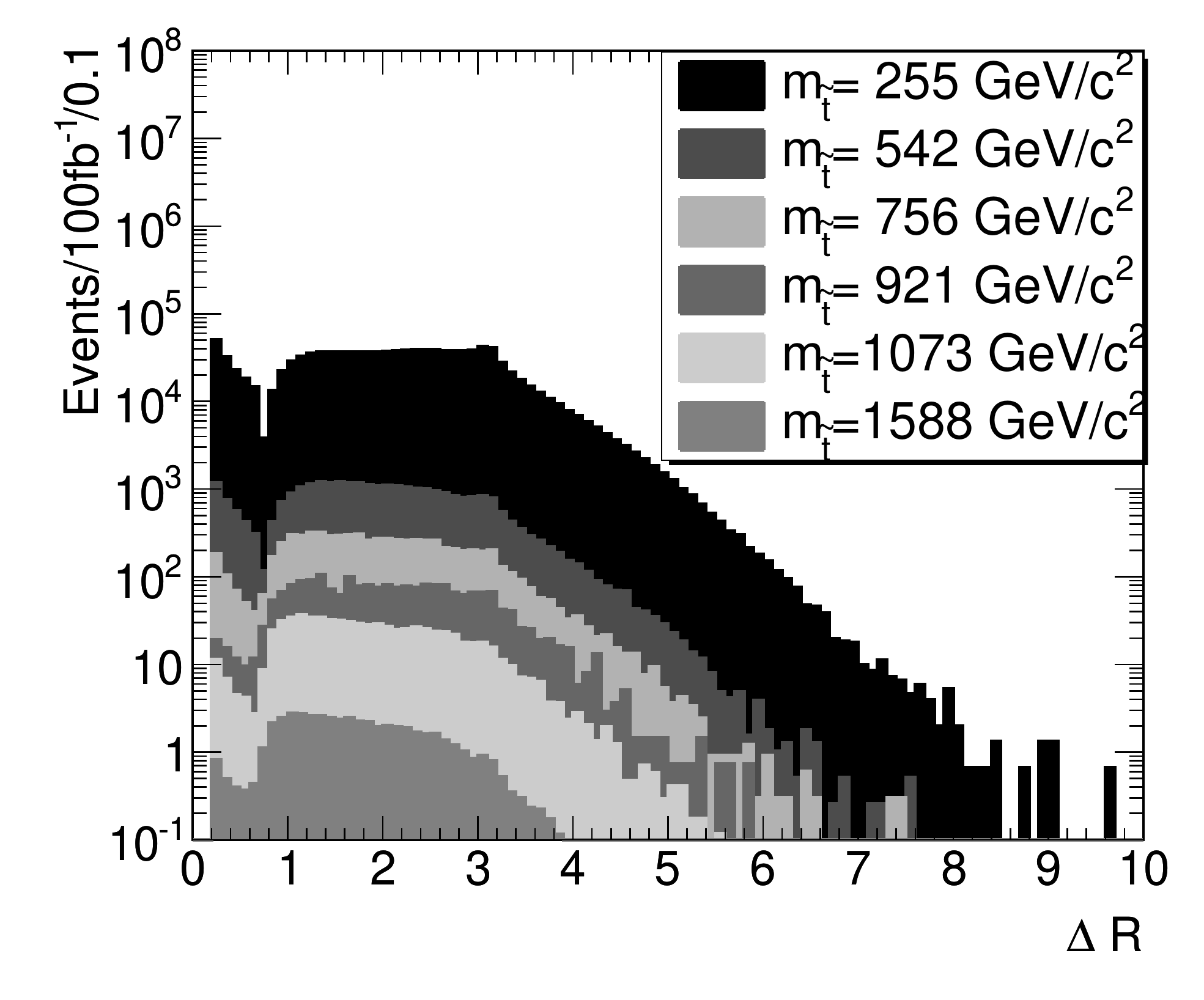}}
\hskip 1 cm
\subfigure[Distance $\Delta R$ between muon and closest jet.]{\label{fig:closestjet_bg}
\includegraphics[width= 0.40\textwidth]{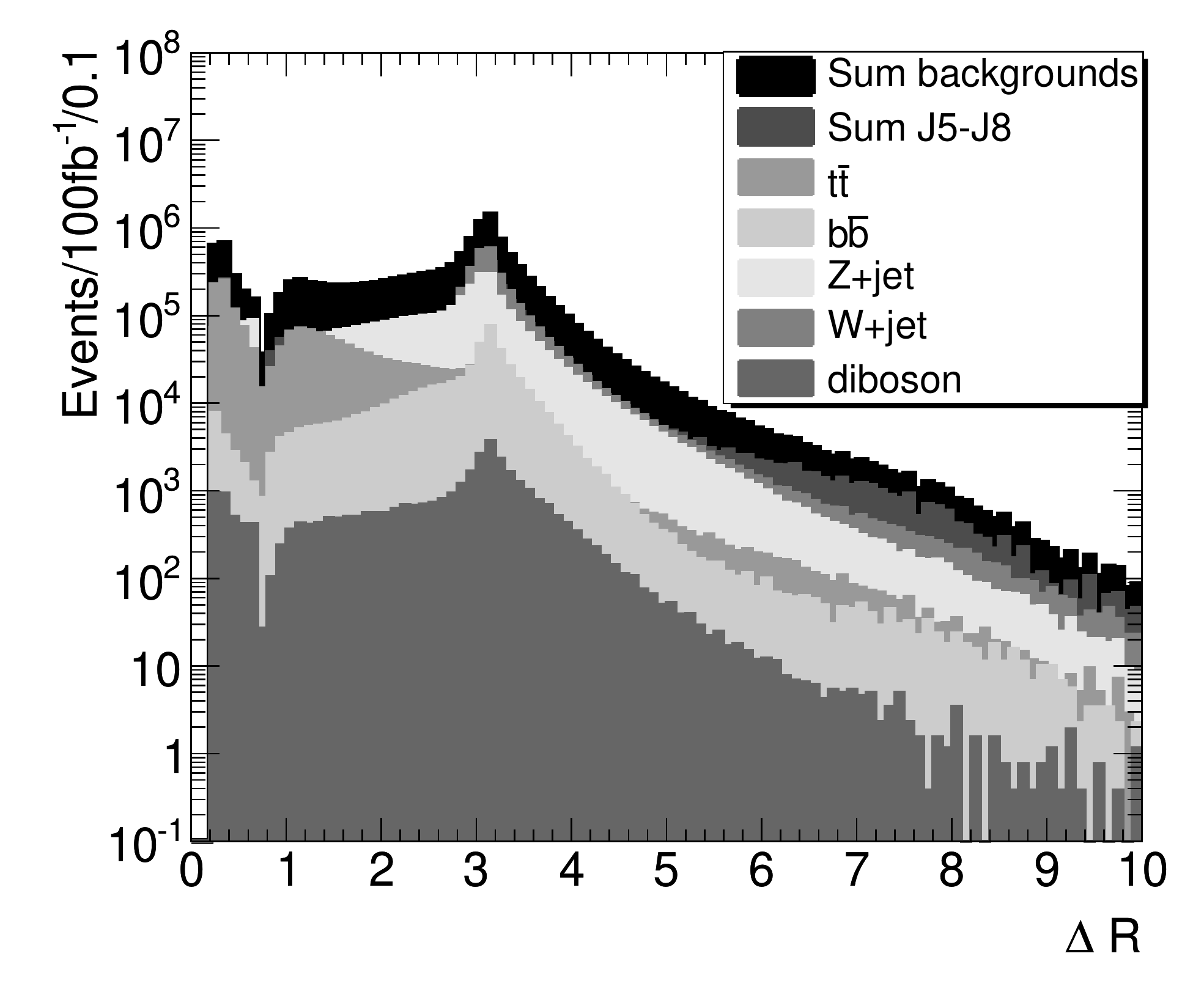} }
\caption{\label{fig:iso} Summed transverse momentum of tracks in a cone $\Delta R <0.2$ and the distance $\Delta R$ between $R$-hadron/muon track and the closest jet for  signal $R$-hadrons \protect \ref{fig:ptcone_signal}, \protect \ref{fig:closestjet_signal} and the various muon backgrounds \protect \ref{fig:ptcone_bg}, \protect \ref{fig:closestjet_bg}.  All plots show the spectrum after a preselection including only events with at least one $R$-hadron/muon that has a transverse momentum larger than 100 $\GeVm$.}

\end{figure}
\begin{table}[ht]\footnotesize
\centering
\begin{tabular}{|c||r||r|r|r|r|r|}
\hline
\multirow{2}{*}{Sample} &Pre- & \multirow{2}{*}{$|\eta|<2.4$} & \multirow{2}{*}{$\beta>0.7$} & $\sum_{\tiny{\Delta R <0.2} }\textrm{p}_{\textrm{T}}$ &\multirow{2}{*}{$\Delta R >0.4$}& \multirow{2}{*}{Combined} \\
 & selection& & & $< 10\, \GeVm$ & &\\ \hline
255&77&97&85&100&96&73\\ \hline
542&93&98&73&100&96&61\\ \hline
756&97&99&65&100&98&54\\ \hline
921&64&99&61&100&99&52\\ \hline
1073&98&99&56&100&98&47\\ \hline
1588&97&99&43&100&98&37\\ \hline
QCD J5&0.17&99&100&12&97&3\\ \hline
QCD J6&0.98&100&100&9&95&2\\ \hline
QCD J7&0.97&100&100&7&93&2\\ \hline
QCD J8&1&100&100&5&90&2\\ \hline
${t}\bar{t}$&2&98&100&93&67&58\\ \hline
${b}\bar{b}$&0.08&97&100&5&99&1\\ \hline
$Z$ + jet&15&89&100&100&95&82\\ \hline
$W$ + jet&4&83&100&98&90&72\\ \hline
$WW$,$WZ$,$ZZ$ &0.14&85&100&99&96&79\\ \hline
\end{tabular}

\caption{\label{tab:effectofCuts}Percentage of remaining events after a given cut. The pre-selection column is the fraction of events remaining after requiring at least one muon with transverse momentum larger than 100 $\GeVm$. The remaining five columns show the efficiency (in percentage points) of each cut separately, relative to the preselected sample. }
\end{table}
\subsection{Data selection and detector acceptance cuts}
A candidate event sample was created by requiring at least one $R$-hadron/muon with transverse momentum larger than 100 $\GeVm$.  The $p_{T}$ spectrum of the $R$-hadrons and muons after this preselection is shown in Figures \ref{fig:pt_signal} and \ref{fig:pt_bg} respectively. The preselection removes a substantial fraction of the background, while most of the signal is kept, which is also seen in Table \ref{tab:effectofCuts}. As expected the  signal retention is larger for high masses. The exception for the 921 $\GeV$ stop mass case is explained  by Table \ref{tab:stopPairProd} where we see that 35\% of the total SUSY production give rise to events where no stop pairs are produced. Out of the events in this sample that contain stops 98\% pass the pre-selection.
\par $R$-hadrons/muons  are required to be within the pseudorapidity range $|\eta|<2.4$. This requirement is motivated by the geometrical acceptance of the ATLAS first level muon trigger system. which extends from $\eta =0$ to $\eta = \pm 2.4$. The signal and background distributions of $\eta$ is shown in Figures \ref{fig:eta_signal} and \ref{fig:eta_bg}. 

\par Since $R$-hadrons are quite heavy they can move with a speed $\beta=\frac{v}{c} <1$. As the particles need to reach the trigger stations within 25 ns to be associated with the correct bunch crossing, the requirement of $\beta >0.7$ is applied. The $\beta$-distributions for signal and background are shown in Figure \ref{fig:beta_signal} and \ref{fig:beta_bg} respectively. Comparison of the two figures shows that the $\beta$-spectrum for the signal tends to peak at lower values for higher masses, while the background sample has nearly all events at $\beta =1$. This is confirmed by Table \ref{tab:effectofCuts} showing that 100\% of events in each background sample pass this cut and that the fraction of signal events passing the cut decreases rapidly with mass. 
\par In addition to the pre-selection and detector acceptance cuts, particle isolation cuts are imposed. Firstly we require that the sum of the transverse momenta of tracks  within a distance $\Delta R = \sqrt{(\Delta \eta)^{2} + (\Delta \phi)^{2}} <0.2$ of the R-hadron/muon track is less than 10 $\GeVm$. Figures \ref{fig:ptcone_signal} and \ref{fig:ptcone_bg} show the summed $p_{T}$ of tracks within a cone of size 0.2 around the $R$-hadron/muon track. The background distribution (Figure \ref{fig:ptcone_bg}) shows that muons are very often produced close to other high $p_{T}$ tracks, while this is to a much lesser extent the case for the signal, $\sim$90\% or more of the QCD background is removed by this cut, cf. Table \ref{tab:effectofCuts}. Most of the muons remaining after these cuts originate from the decay of $W$ or $Z$ bosons.
\par The other isolation criteria requires that the distance $\Delta R$ between the muon and the closest jet is larger than 0.4. This result is highly dependent on the jet-algorithm used; in this case jets are defined using a cone algorithm with cone size $=0.7$ and a 50 GeV energy threshold. The distance between a $R$-hadron/muon and the closest jet is plotted in Figures \ref{fig:closestjet_signal} and \ref{fig:closestjet_bg} respectively. Both distributions show a relatively  good separation between $R$-hadrons/muons and jets, which is also reflected in the second last column of Table \ref{tab:effectofCuts} which shows that more than 90\% of the events are kept for both signal and background.

\par The last column of Table \ref{tab:effectofCuts} shows the total remaining percentage of the signal and background samples after applying all  cuts. The cuts remove between 30\% and 60\% of the signal. The requirement of $\beta > 0.7$ is the one which removes, by far, the most signal.  Out of the background events up to 99\% of the QCD and $b\bar{b}$ samples are rejected by the combined cuts and it is the muon isolation criterium that removes most events. In the case of electroweak processes ( $Z$+jet, $W$+jet and $WW$,$ZW$,$ZZ$ ) the samples are predominantly reduced by the cut on $|\eta|$ and the distance between the muon and the closest jet. The latter requirement also removes the bulk of the rejected  $t\bar{t}$ events.

\begin{figure}[tbp]
\subfigure[Signal]{\label{fig:leadingjet_signal}
\includegraphics[width= 0.45\textwidth, height=5.5 cm]{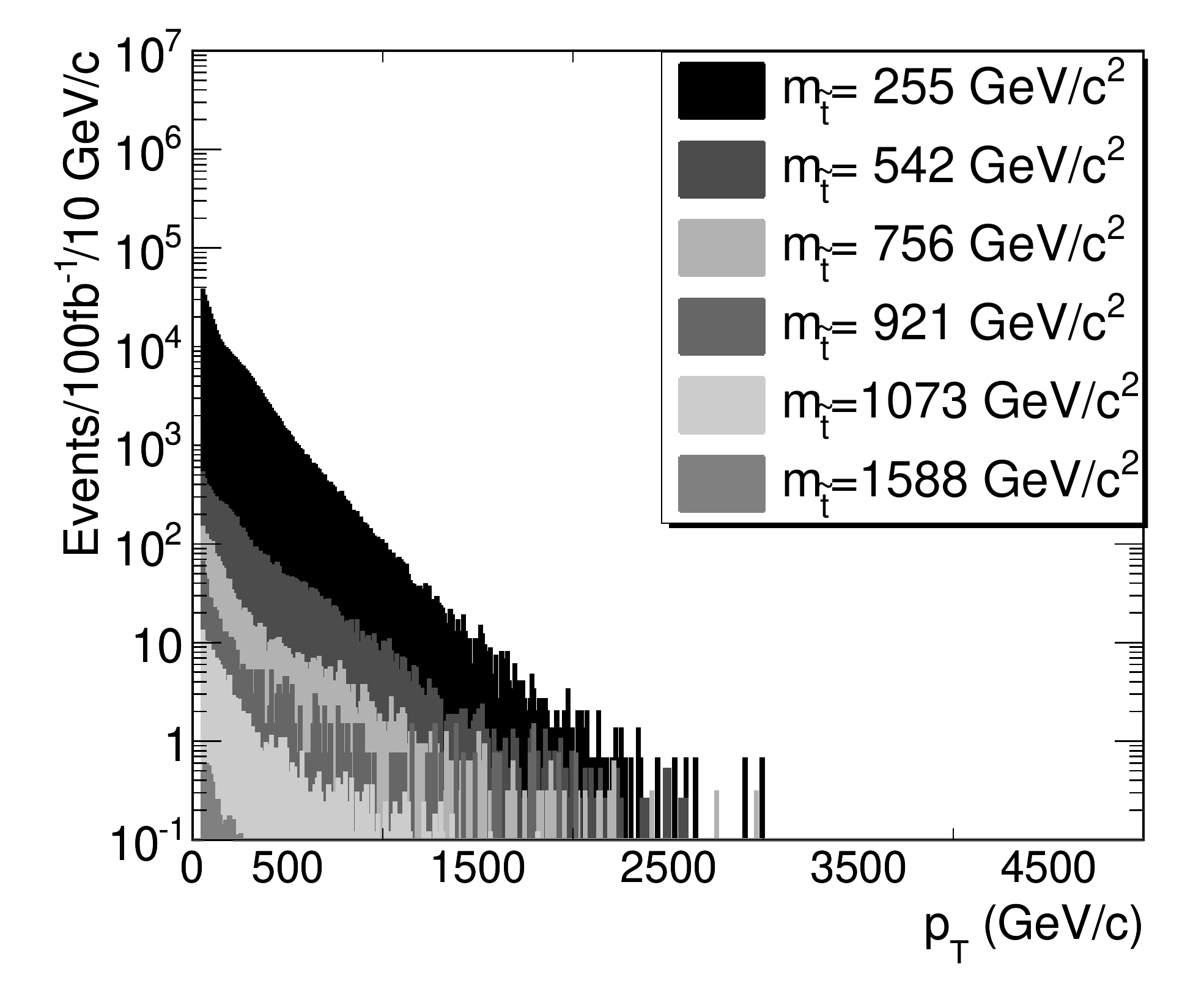} }
\subfigure[Background.]{\label{fig:leadingjet_bg}
\includegraphics[width= 0.45\textwidth, height=5.5 cm]{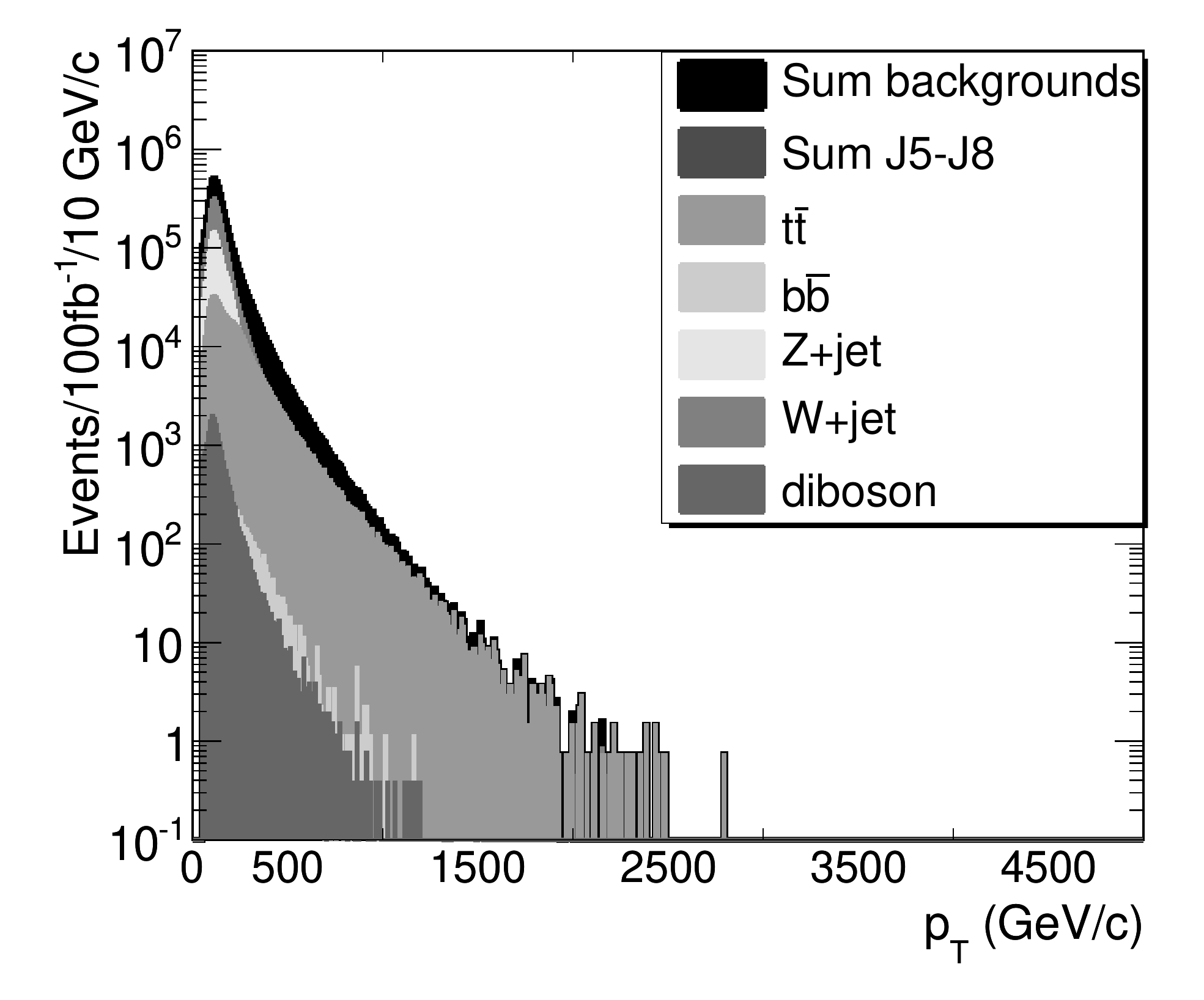} }
\caption{Leading jet transverse momentum distribution for signal \protect \ref{fig:leadingjet_signal} and background \protect \ref{fig:leadingjet_bg}. }
\end{figure}

\begin{table}\footnotesize
\centering
\begin{tabular}{|c|r|r|r|r|}
\hline
\multirow{2}{*}{Sample} & \multicolumn{4}{c|}{Number of Events} \\ \cline{2-5}
& Pre- & Isolation and & Jet $p_{T}$& Jet and $\mu$ \\ 
& selection&Acceptance &$<100$ $\GeVm$ &$p_{T}$-cut \\
\hline
255&1.02885e+06&753560&485212&18234\\ \hline
542&24019&14578&7702&2582\\ \hline
756&5946&3223&1860&1083\\ \hline
921&1890&978&643&438\\ \hline
1073&582&272&156&127\\ \hline
1588&33&12&6&5\\ \hline
j5&2.17032e+06&62999&4734&0\\ \hline
j6&353081&8100&127&0\\ \hline
j7&5556&97&1&0\\ \hline
j8&3&0.04&0&0\\ \hline
$\textrm{t}\bar{\textrm{t}}$&1.07456e+06&626064&105812&24\\ \hline
$\textrm{b}\bar{\textrm{b}}$&450008&5971&910&0\\ \hline
Z + jet&2.03223e+06&1.66846e+06&518456&99\\ \hline
W + jet&4.68819e+06&3.35225e+06&992362&34\\ \hline
WW,WZ,ZZ&32469&25701&12586&18\\ \hline
\end{tabular}

\caption{\label{tab:ptCuts} Number of events remaining after preselection cut, isolation and acceptance cuts, leading jet cut and leading jet + leading muon cut. }
\end{table}

%%%%%%%%%%%%%%%%%%%%%%%%%%%%%%%%%%%%%%%%%%%%%%%%%%
%****************************SIGNAL ENHANCEMENT*********************************%
%%%%%%%%%%%%%%%%%%%%%%%%%%%%%%%%%%%%%%%%%%%%%%%%%%
\subsection{Enhancing the stop $R$-hadron signal}
After the acceptance cuts and the isolation requirements on the muon
candidates we are left with from 12 to order of 750 000 events for
the signal samples, while the total number
of background events is over 5 million events. The  $W/Z$ + jet events
are the dominant background followed by QCD and $t\bar{t}$ events, as shown in Table  \ref{tab:ptCuts}.

\begin{figure}
\centering
\includegraphics[width= 0.55\textwidth]{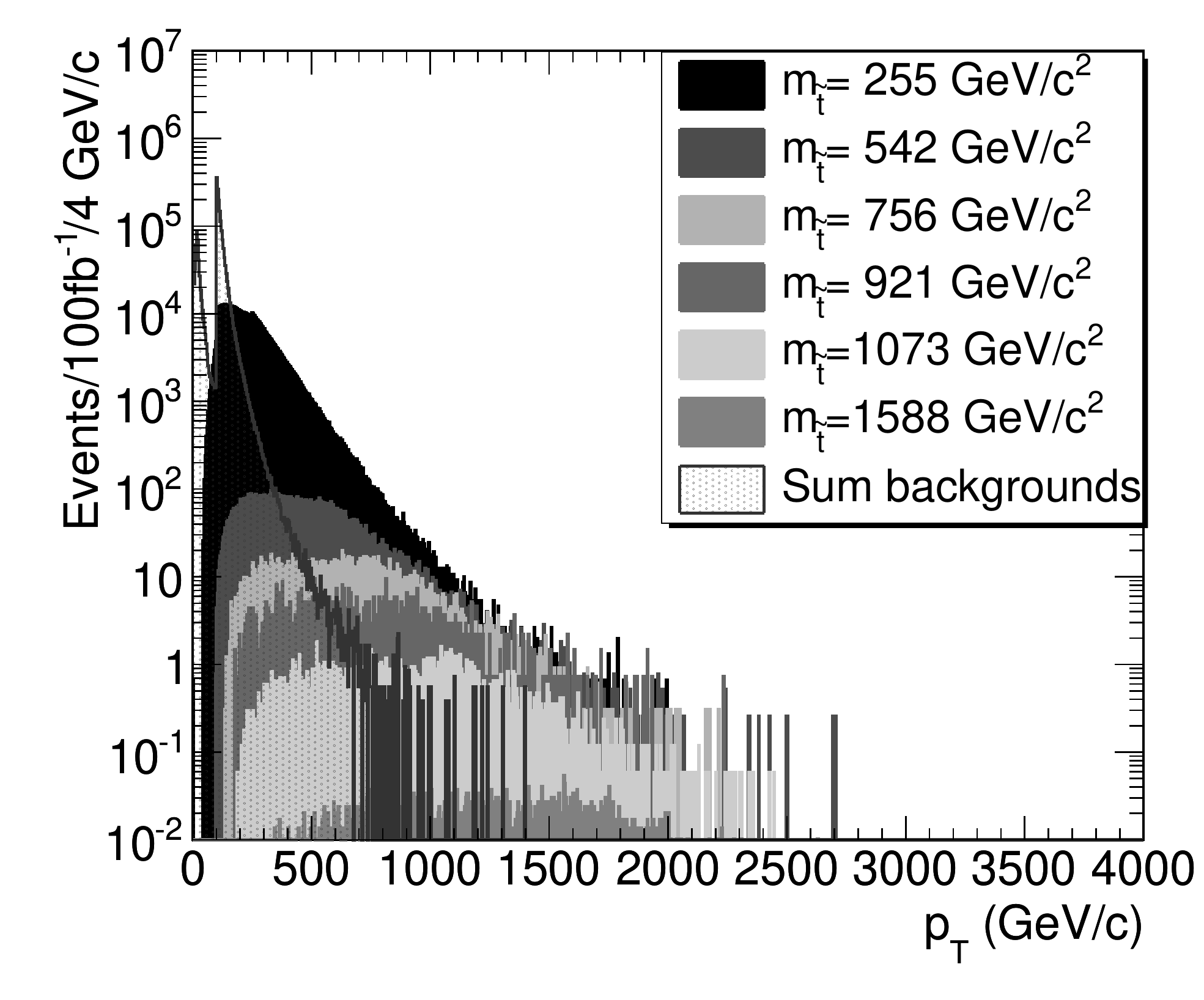}
\caption{\label{fig:lowJetCut} Transverse momentum distribution for all $R$-hadron signal samples and the summed muon background  after applying preselection, detector acceptance and isolation cuts and requiring that  $p_{T}<100\, \GeVm$ for the leading jet.}

\end{figure}
\begin{figure}[ht]
\subfigure[$m_{\tilde{t}_{1}}=255\, \GeV$.]{\label{fig:sig_255}
\includegraphics[width= 0.40\textwidth]{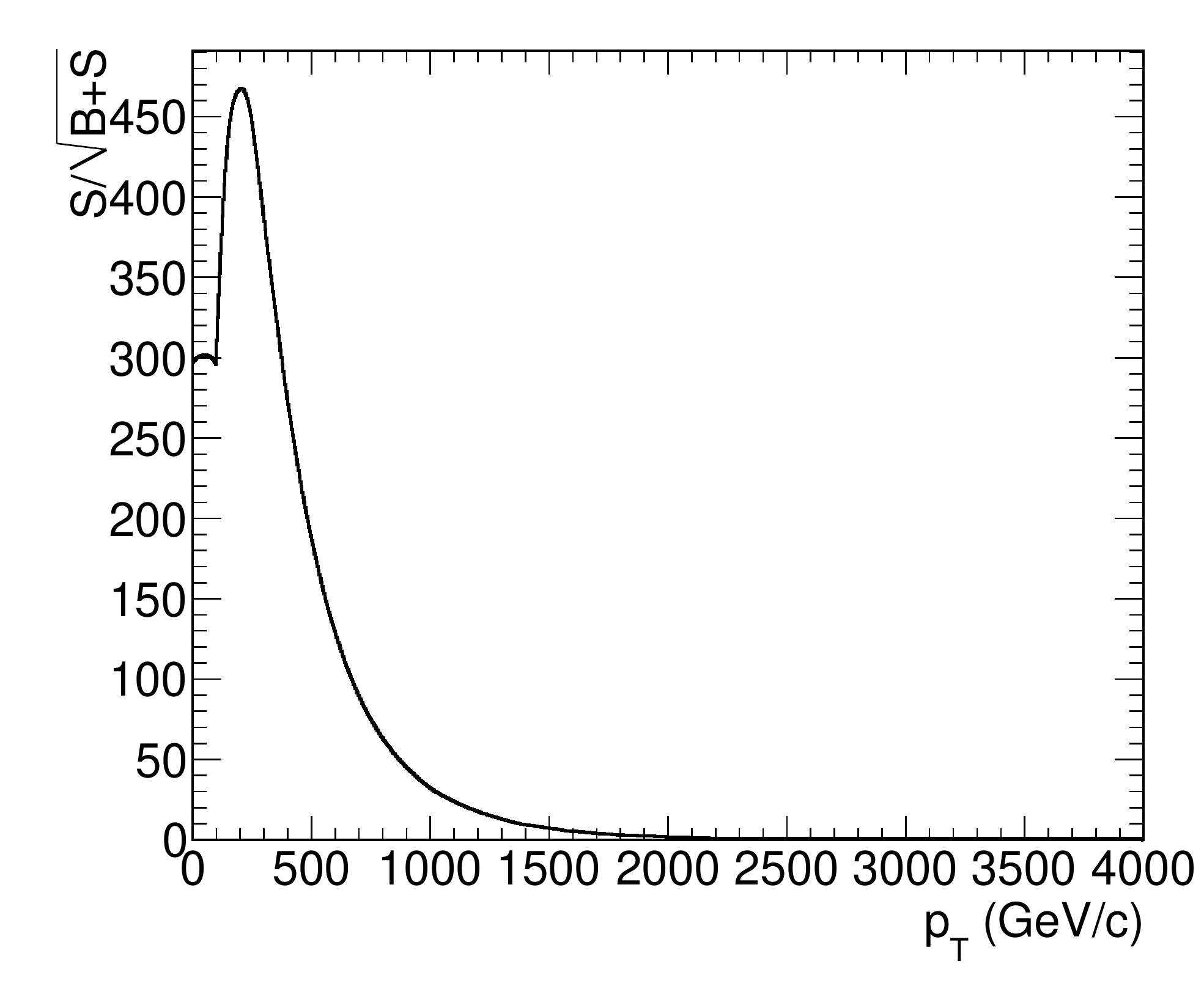} }
\subfigure[$m_{\tilde{t}_{1}}=542\, \GeV$.]{\label{fig:sig_542}
\includegraphics[width= 0.40\textwidth]{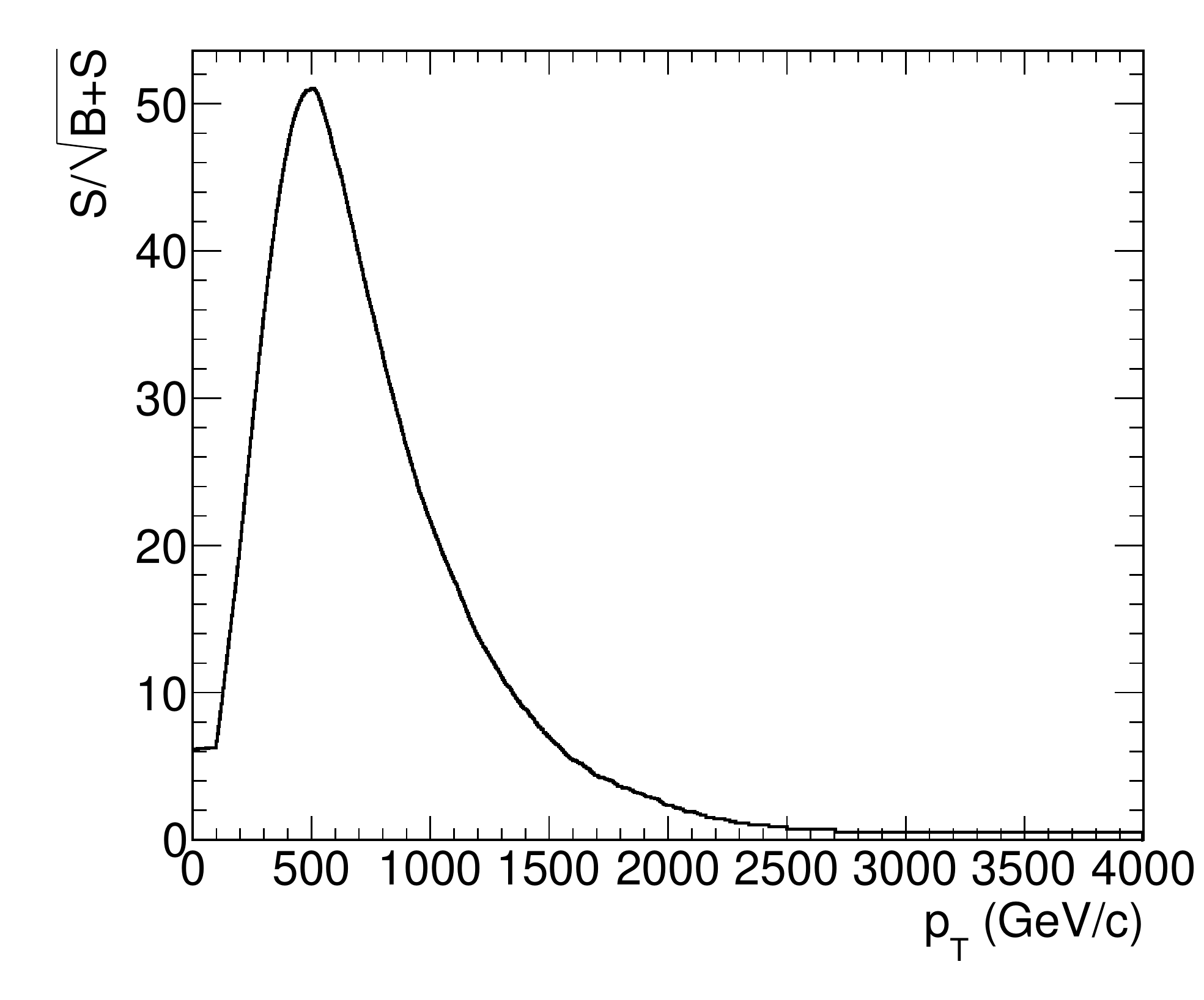} }

\subfigure[$m_{\tilde{t}_{1}}=756\, \GeV$.]{\label{fig:sig_756}
\includegraphics[width= 0.40\textwidth]{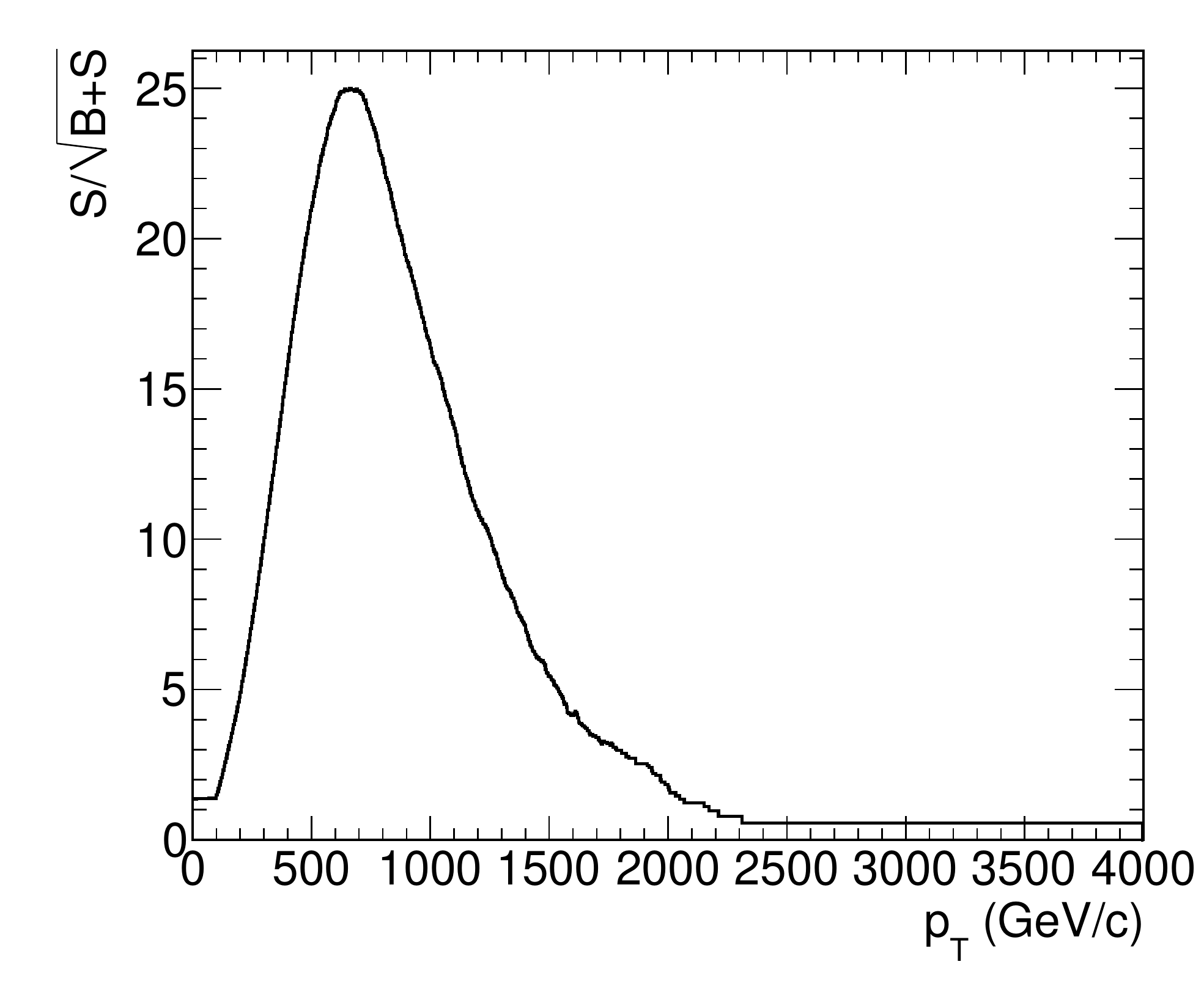}}
\subfigure[$m_{\tilde{t}_{1}}=921\, \GeV$.]{\label{fig:sig_921}
\includegraphics[width= 0.40\textwidth]{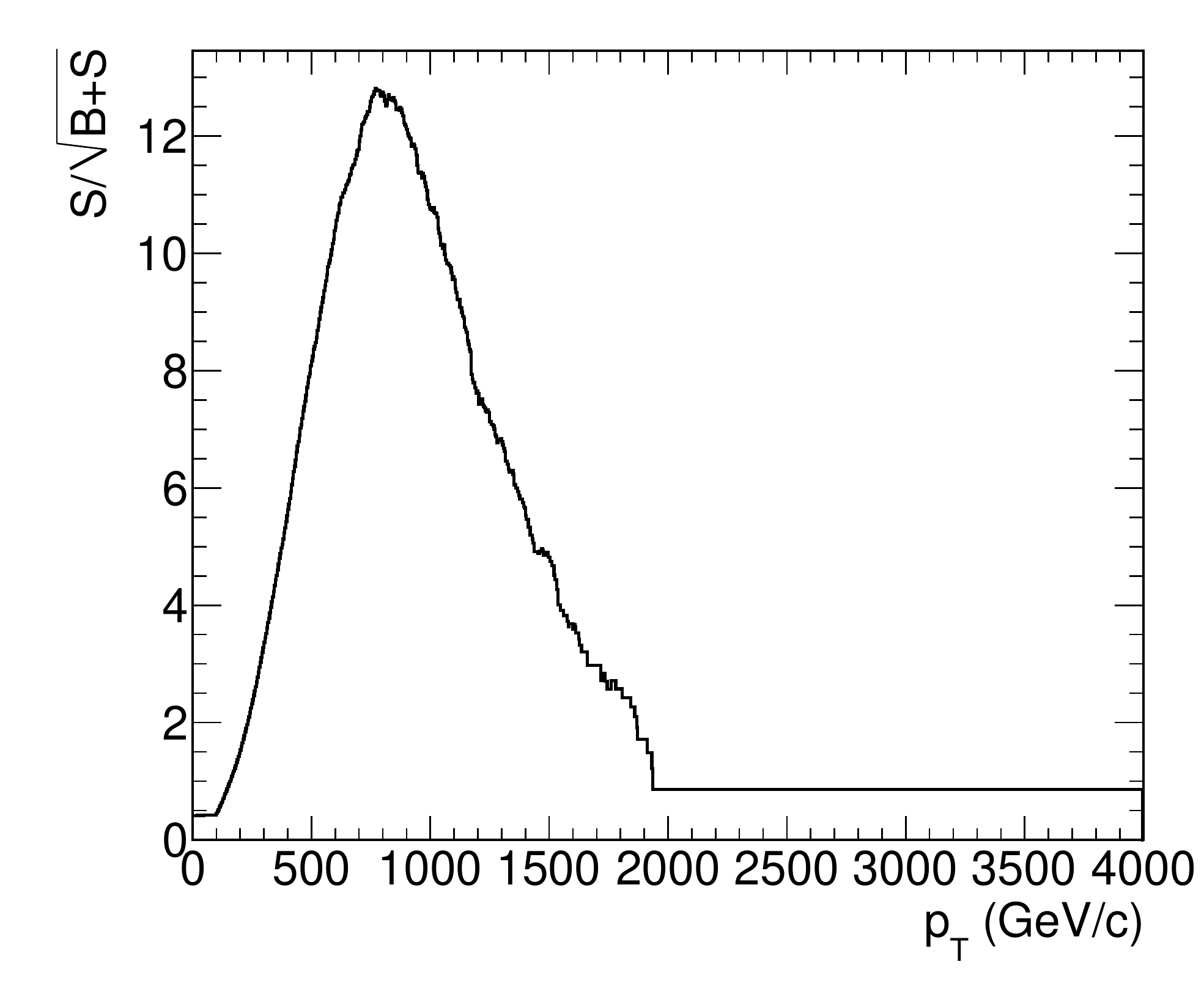} }

\subfigure[$m_{\tilde{t}_{1}}=1073\, \GeV$.]{\label{fig:sig_1073}
\includegraphics[width= 0.40\textwidth]{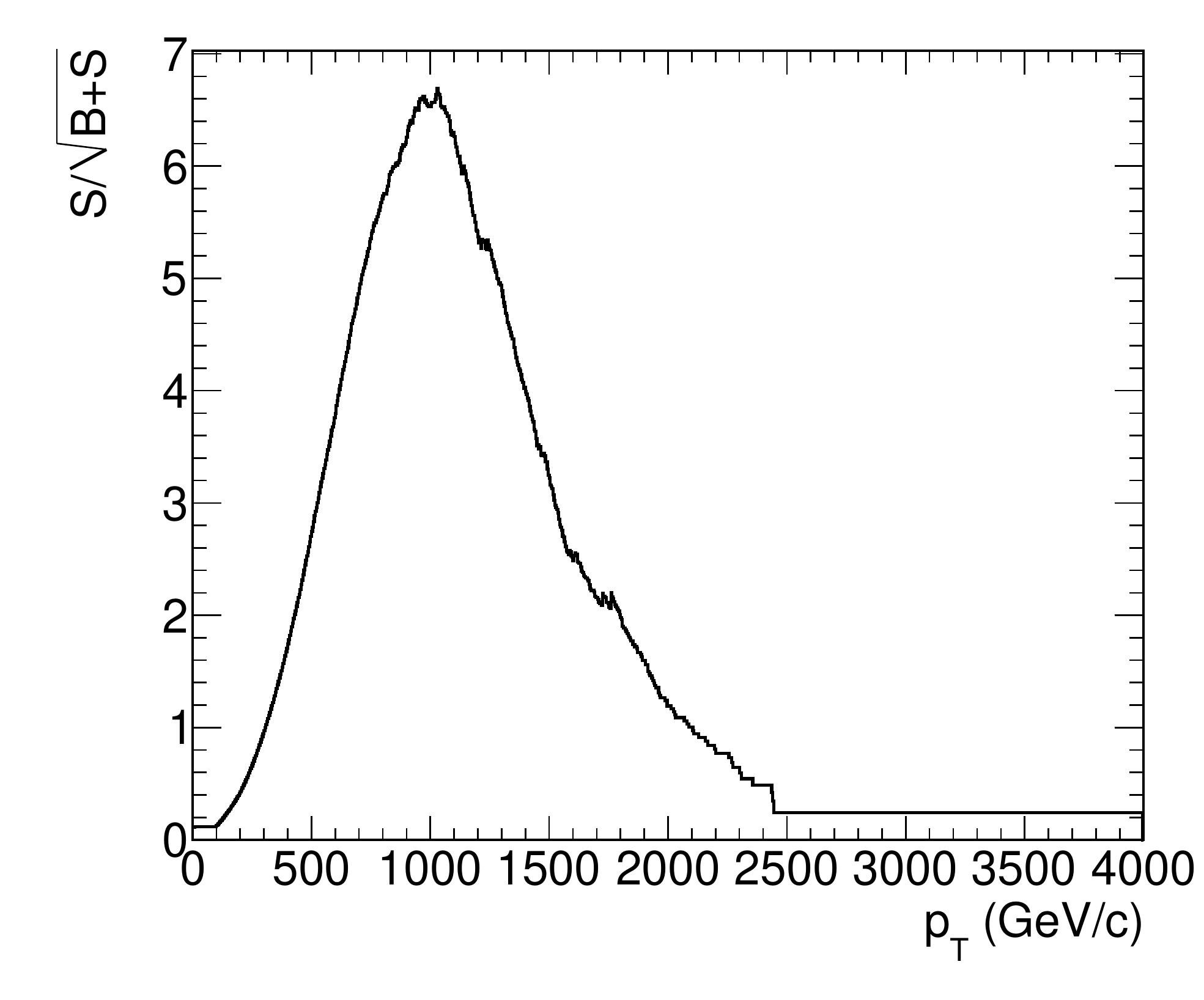}}
\subfigure[$m_{\tilde{t}_{1}}=1588\, \GeV$.]{\label{fig:sig_1588}
\includegraphics[width= 0.40\textwidth]{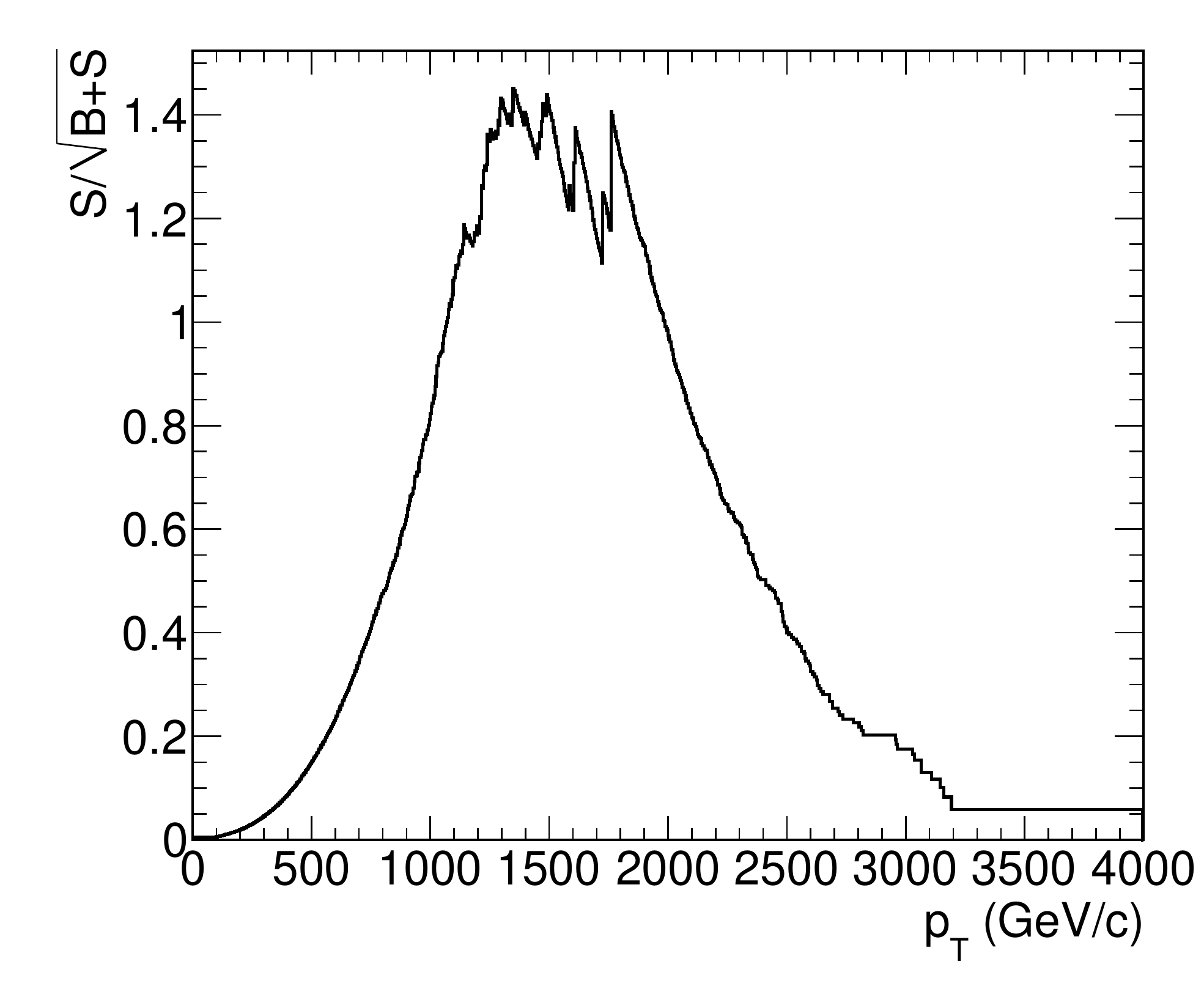} }
\caption{\label{fig:significance} $S/\sqrt{B+S}$ for all signal samples as a function of the cut on the leading muon transverse momentum.}
\end{figure}

After the preselection requirement of a transverse momentum on the muon candidate $( p_{T}(\mu) > 100\, \GeVm)$ the main contribution from the electroweak processes comes
from  bosons  produced in 2 \( \rightarrow \) 2
processes in which a jet recoils against the boson. This gives a boost
to  the boson which leads to high \( p_{T} \) of the muons
produced when it decays.

In the signal samples jets can be produced in various ways, in
the case of direct stop pair-production jets originate from initial
and final state radiation as well as the underlying events.
 When stop pairs are indirectly produced through the decay of heavier squarks,
jets originate from hadronic decays of $W$ or $Z$ bosons produced in, for
instance, sbottom decay ($\tilde{b}_{i} \rightarrow \tilde{t}_{1} + W$)
or stop decay ($\tilde{t}_2 \rightarrow \tilde{t}_{1} + Z$).

In the models considered here however, the mass-splitting between the
third generation squarks is relatively small, leading to rather low
momentum for the accompanying jets in these decay chains.
The distribution of the momentum of the leading jet
in the signal and background samples, shown in Figures
\ref{fig:leadingjet_signal} and \ref{fig:leadingjet_bg}
demonstrates the expected difference between signal and
background.

To exploit this difference to further enhance the signal we require
the leading jet to have transverse momentum below 100 $\GeVm$
which leaves the number of events shown in column 4 of Table \ref{tab:ptCuts}.
The $W$ and $Z$ samples are reduced by 70\%, while the corresponding
numbers for the signal vary  between 40\% and 50 \%.  Also note
that both the $t\bar{t}$ and $b\bar{b}$ samples as well as the QCD samples are
significantly reduced.

Figure \ref{fig:lowJetCut} shows the transverse momentum distribution for all signal  samples and the summed muon background  after applying the cut on the $p_{T}$ of the leading jet. The remaining background decreases rapidly from approximately 100 $\GeVm$ and motivates an additional hard $p_{T}$-cut.

\par To further reduce the remaining background  a cut on the leading $R$-hadron/muon is imposed. We plot $S/\sqrt{B+S}$ as a function of the cut on muon-$p_{T}$ to find the cut value that optimises the signal to background ratio, see Figure \ref{fig:significance}. Based on these plots a requirement of $p_{T}>500$ GeV for the leading muon is imposed. The remaining events are listed in the last column of  Table \ref{tab:ptCuts} and from these numbers we can easily calculate the signal significance $S/\sqrt{S+B}$ which is larger than 5 for all signal samples except the heaviest mass point at 1588 $\GeV$.

\section{Conclusion}
In this work we have studied the possibility for a long-lived stop to arise in supersymmetric models with a neutralino LSP. This is achieved when the mass difference between the stop and neutralino is extremely small, less than the $c$-quark mass. The small mass difference increases the coannihilation cross section leaving the neutralino relic density smaller than the WMAP-result. This implies either that the neutralinos coming from these models can only constitute a part of the total dark matter and that there in addition must be other particles/mechanisms making up rest, or that the lower bound on the relic density, as derived from the WMAP data, is invalid due to modifications to the expansion rate of the universe prior to Big Bang nucleosynthesis as proposed e.g. in Ref. \cite{nazila}.
\par
We have also seen that SUSY models with a long-lived stop give rise to non-traditional SUSY signatures. In many of the studied models the main processes will be various kinds of Higgs production or stop-pair-production, none of which give rise to the standard leptons plus jets plus missing transverse energy signatures. If such models are realised in nature the discovery of a long-lived stop could be the first sign of physics beyond the standard model.
\par
Finally we have studied six SUSY models where stop pair-production dominates. By using a range of final state observables, it should be possible to discover stable stops with masses up to around 1 $\TeV$ at the LHC. Should timing-specific observables, such as the reconstructed speed of a $R$-hadron candidate, be used in a search, the mass reach could be higher still.
\newline
\newline
{\bf Acknowledgements:} We want to thank M. M\"uhlleitner and P. Skands for help concerning the programs {\sc Sdecay} and {\sc Pythia} respectively, A. Raklev for helpful supersymmetry discussions and  A. Sellerholm for taking part in the initial steps of this work. 

We are grateful to the Swedish Research Council (VR) for support. D. Milstead is a Royal Swedish Academy Research Fellow supported by a grant from the Knut and Alice Wallenberg Foundation.

\bibliographystyle{JHEP}

\bibliography{%
stablestop_vJHEP}
\clearpage

\end{document}